\documentclass[pre,twocolumn]{revtex4-2}

\usepackage{graphicx}% Include figure files
\usepackage{dcolumn}% Align table columns on decimal point
\usepackage{bm}% bold math
\usepackage[utf8]{inputenc}
\usepackage[T1]{fontenc}
\usepackage{mathptmx}

\usepackage{amsmath}
\usepackage{mathtools}
\usepackage{amssymb}

\usepackage{pst-node}

\usepackage{ulem}
\usepackage{mhchem}
\usepackage{wasysym}
\usepackage{tikz}
\usetikzlibrary{decorations.pathmorphing,patterns}
\usepackage{pst-coil}
\usepackage{lipsum} 
\usetikzlibrary{graphs}

\usepackage{listings}
\usepackage{color}
\definecolor{dkgreen}{rgb}{0,0.6,0}
\definecolor{gray}{rgb}{0.5,0.5,0.5}
\definecolor{mauve}{rgb}{0.58,0,0.82}

\usepackage{verbatim}   % useful for program listings
\usepackage{color}      % use if color is used in text
\usepackage{subfigure}  % use for side-by-side figures
\usepackage{hyperref}   % use for hypertext links, including those to external documents and URLs

\usepackage{braket} 

\newcommand{\bnabla}{\mbox{\boldmath $\nabla$}}

\newcommand{\ba}{\begin{eqnarray}}
\newcommand{\ea}{\end{eqnarray}}
\newcommand{\be}{\begin{equation}}
\newcommand{\ee}{\end{equation}}

\pdfoutput=1

\begin{document}

%==============================================================
\title{Run-and-tumble particles in slit geometry as a splitting probability problem}
%\title{Unhappiness of run-and-tumble particles in slit confinement}
%==============================================================

\author{Derek Frydel}
\affiliation{Department of Chemistry, Universidad Técnica Federico Santa María, Campus San Joaquin, Santiago, Chile}

\date{\today}

\begin{abstract}

Run-and-tumble particles confined between two walls seem like a simple enough problem to possess 
analytical tractability. Yet up to date no satisfactory analysis is available for dimensions higher than one. 
This work contributes to the theoretical understanding of this system by reinterpreting it as a splitting 
probability problem. Such reinterpretation permits us to formulate the problem as the integral equation, 
rather than a more standard differential equation based on the Fokker-Planck equation.  In addition to 
providing an analogy with another phenomenon, the reinterpretation permits a new type of analysis, 
yields useful results, and offers some analytical tractability.

\end{abstract}

\pacs{
}

\maketitle
%------------------------------------------------

\section{Introduction}

The opening sentence of Anna Karenina states:  "All happy families are alike; 
each unhappy family is unhappy in its own way."   This sounds a bit like statistical mechanics 
if we substitute happiness with equilibrium and unhappiness with non-equilibrium.  
%This assertion may appear like an attempt to characterize an experience of 
%working with systems in non-equilibrium.  While systems in equilibrium are similar in a sense 
%that their distributions are described by a Boltzmann weight, stationary systems that are 
%out-of-equilibrium cannot be described by a general statistical weight.  
Each non-equilibrium system "is unhappy in its own way."  
%This is why general statements and 
%relations regarding non-equilibrium are scarce and in demand.   

Anyone working even with the most idealized non-equilibrium systems might have recognized that a 
slight modification can bring about significant changes.  Not just changes of physical properties 
but changes that 
%call for the use of 
require 
very different tools of analysis.  For example, solving the problem of active particles 
in a harmonic trap is not particularly helpful when trying to analyze the system of particles confined 
between two walls.  Changing interest from one system to another feels like 
%It is as if 
one has to start the learning process all over again.  This is frustrating but also 
makes non-equilibrium statistical mechanics interesting.

Run-and-tumble particle model (RTP) is probably the most ideal system of active particles and it could 
be regarded as an ideal-gas-model of active matter.  It has all the features of an active particle 
implemented in the most straightforward way.  Yet even for the simplest of confinements, the resulting 
theoretical problem can be overwhelming.

A typical starting point of theoretical analysis is the Fokker-Planck equation, which in most cases 
cannot be solved exactly.  One can assume that by formulating the problem in terms of a marginal 
stationary distribution with a reduced number of variables 
%(instead of a full distribution in the Fokker-Planck equation) 
some complexity can be reduced.  The difficulty with 
this idea is that it is not clear how to arrive at such a formulation.  There is no direct path from the 
Fokker-Planck equation to such an alternative formulation.  
%There is no general technique that reduces one framework to another.   

%The lack of standard techniques forces us to be more adventurous and unorthodox when choosing methodology.  

For RTP particles in a harmonic trap in two-dimensions it was possible to obtain a third order-differential 
equation for a stationary marginal distribution \cite{PRE-106-Frydel-2022}
%not from the Fokker-Planck equation but 
by analyzing the moments of a marginal distribution \cite{PRL-Caraglio-2022,POF-Frydel-2023}.  
The same trick no longer works when applied to slit geometry.  One of the features of 
slit geometry is that a fraction of particles remains adsorbed onto a wall.  This 
leads to singularities at the location of a wall, making mathematics complicated.  
Thus, despite its simplicity, the system of RTP particles between two walls in a stationary state 
remains an unsolved problem.  

This article is an attempt to rectify this situation.  
To do this, we reinterpret RTP particles between two walls in a stationary state as the splitting 
probability problem \cite{PRL-Klinger-2022,Ziff-JSP-2006,kampen-1992,redner-2001,hughes-1995}, 
where the splitting  probability is the probability that a random walker, initially 
at some location in the region between the two walls, reaches the wall on the right-hand-side without 
reaching first the wall on the left-hand-side.  
We arrive at this alternative description by modifying the microscopic motion of
particles:  rather than using a continuous time Langevin dynamics, our particles move by discontinuous
jumps.  Although different on the microscopic level, on the macroscopic level the two systems are 
identical.  This permits us to choose the macroscopic framework that is more convenient ---
in this case the framework based on discontinuous jumps and represented by a integral equation.  

%problem \cite{PRL-Klinger-2022,Ziff-JSP-2006,kampen-1992,redner-2001,hughes-1995}, which permits us
%to reinterpret RTP particles between two walls as the splitting probability problem.  
%The splitting  probability is the probability that a one-dimensional random walker, initially 
%at some location in the region between the two walls, reaches the wall on the right-hand-side without 
%reaching the wall on the left-hand-side first.  
%Consequently, the splitting probability is different from the first arrival time event.  
Other phenomena that exhibit the splitting probability behavior
are the melting probability of a heteropolymer \cite{EPL-Oshanin-2009}, the fixation probability 
of a mutant in population dynamics \cite{moran-1962}, and transmission probability of photons or neutrons 
through a slab of a scattering medium \cite{PRE-Vezzani-2014,PRE-Kaiser-2014,PRE-Robin-2021}

%reinterpreting the system of 
%RTP particles in slit geometry as the splitting probability problem.  This was done by 
%modifying the microscopic process in which particles sample different configurations, from Langevin
%continuous time dynamics to discrete jumps.  The two microscopic processes yield the same 
%stationary distributions if the probability distribution of jump function if correctly selected.  
%This modification of a microscopic processes resulted in a different macroscopic theoretical 
%description, and rather than formulating the problems as a Fokker-Planck differential equation, 
%it was possible to formulate it as an integral equation for the splitting probability.  

The RTP model was originally conceived to represent the motion of bacteria 
\cite{berg-1983,COM-Grognot-2021}.  RTP particle moves at a constant 
velocity but randomly changing orientation.  Orientations change at time intervals
drawn from an exponential distribution.  For a system in one-dimension, there 
are two possible orientations, which simplifies mathematics and allows for 
exact results.   
An extensive analysis of the RTP model in one-dimension
has been done in \cite{Malakar-2018}.  The entropy production rate of the one-dimensional 
system was studied in \cite{PRE-Razin-2020,PRE-Fyrdel-2022,PRE-Gupta-2023}.  
Dynamic properties, including first passage, survival probability, and local time were 
studied in \cite{EPJE-Angelani-2014,PRL-Schehr-2020,JSTAT-Scher-2021,PRE-Kundu-2021}.  
Extensions of the RTP model in one-dimension to include
more than two swimming velocities have been considered in \cite{JPA-Basu-2020,JSTAT-Frydel-2021,POF-Frydel-2022,EPJE-Hartmut-2022}.  
RTP particles in a harmonic potential in one-dimension have been investigated in 
\cite{PRL-Cates-2008,EPL-Cates-2009,JPA-Basu-2020,PRE-106-Frydel-2022,PRE-Scher-2022,PRE-Smith-2022,Smith-PRE-2022a,POF-Frydel-2023}.  
RTP particles in other types of potentials and under different conditions has been considered in 
\cite{PRE-Dhar-2019, PRE-Farago-2024, PRE-connor-2023,PRE-Dhar-2019,EPL-Doussal-2020,JSTAT-Wijland-2023,JPA-Angelani-2015,JSTAT-Bressloff-2023,PRL-Tailleur-2019,EPL-Detcheverry-2015,PRE-Dean-2021,JPA-Angelani-2017}.  Recently, effective diffusion of RTP particles in a slit geometry
has been investigated in \cite{PRR-RTP-2024}.

\section{The model}

The run-and-tumble dynamics consists of two stages.  During the "run" stage a particle moves with 
the swimming velocity ${\bf v} = v_0{\bf u}$, where $v_0$ is the constant magnitude and ${\bf u}$ is the unit vector 
designating the direction of motion.  During the "tumble" stage, which occurs instantaneously, 
the unit vector ${\bf u}$ changes to any other orientation with uniform probability.  The
time $t_p$ during which a velocity persists in a given orientation is drawn from an exponential distribution 
\be
p_t = \frac{ e^{-t_p / \tau}} {\tau}, 
\label{eq:pt}
\ee 
where $\tau = \langle t\rangle$ is the average persistence time during which a swimming velocity persists 
in a given direction.

In the absence of an external potential, the Fokker-Planck equation for RTP particles 
(for convenience given in 2D) is 
\be
\dot n   =   -v_0 {\bf u} \cdot \bnabla n   -   \frac{ 1 }{\tau} \left(  n -   \int_0^{2\pi} \frac{d\theta}{2\pi}\, n \right), 
\label{eq:FP2D}
\ee
where $n \equiv n({\bf r},\theta,t)$ is the time dependent distribution.  
Orientation of a swimming motion in 2D is given by ${\bf u} = (\cos\theta,\sin\theta)$, 
where it depends on an angle $\theta\in [0,2\pi]$.  
For particles confined between two parallel walls at $x=0$ and $x=L$, the stationary state 
is governed by a one-dimensional Fokker-Planck equation, 
\be
0   =   -v_0\cos\theta n'   -   \frac{ 1 }{\tau} \left(  n -   \int_0^{2\pi} \frac{d\theta}{2\pi}\, n \right),
\label{eq:FP2D-1D}
\ee
where $v_0\cos\theta$ is the projection of a velocity vector on the $x$-axis.
%and $n\equiv n(x,\theta)$.  
The above equation can be written as 
%Eq. (\ref{eq:FP2D-1D}) and rewrite it for the distribution $n\equiv n(x,v)$ as 
\be
0   =   -v n'   -    \frac{ 1 }{\tau} \left(  n  -   \int_{-v_0}^{v_0} dv\, p_v n\right),
\label{eq:FP}
\ee
where $n\equiv n(x,v)$, $v\equiv v_0 \cos\theta$ is the projection of the swimming
velocity onto the $x$-axis (the direction perpendicular
to the walls) and $p_v(v)$ is the probability distribution of those velocities.
% defined on the interval $v\in[-v_0,v_0]$.  
The interaction with the walls is incorporated via the boundary conditions 
$0 = v n(0,v) = v n(L,v)$, implying zero flux at the walls.

The advantage of Eq. (\ref{eq:FP}) is that it applies to any dimension $d$, 
unlike Eq. (\ref{eq:FP2D-1D}) that is specific to $d=2$.  The dependence on $d$ enters 
through the probability distribution $p_v$,  
\be
p_v
=  \frac{1}{2}
\begin{cases}
     \delta(v-v_0) + \delta(v+v_0), & \text{for $d=1$},\\
     \frac{2}{\pi} \frac{1}{\sqrt{v_0^2-v^2}}, & \text{for $d=2$}, \\
      \frac{1}{v_0}, & \text{for $d=3$}, \\
  \end{cases}
\label{eq:pv}
\ee
For $d=1$, two possible swimming directions are $v=\pm v_0$ \cite{PRE-Razin-2020}.  
For $d=2$ and $d=3$, $p_v$ is calculated using the variable change $v = v_0 \cos\theta$, 
%that modifies the integral over angular orientation as 
$$
\int_0^{2\pi} d\theta\, ~\to~  \int_{-v_0}^{v_0} dv\,\frac{1}{\sqrt{v_0^2-v^2}}, 
$$
and 
$$
\int_0^{2\pi} d\phi \int_0^{\pi} d\theta\, \sin\theta   ~\to~  \int_{-v_0}^{v_0} dv.  
$$
Within the formalism of Eq. (\ref{eq:FP}), the marginal distribution is defined as 
\be
\rho(x) = \int_{-v_0}^{v_0} dv\, p_v(v) n(x,v).  
\ee

One difficulty in dealing with Eq. (\ref{eq:FP}) comes from the interaction with the walls, which leads to
"dynamic" adsorption of particles.  It is not an adsorption in a usual sense (since walls do not have 
any attractive potential), but a result of a combined effect of the persistence time and overdamped 
dynamics.  Once an active particle comes in contact with a wall, it does not bounce from it but 
continues pushing against it, becoming effectively adsorbed and exerting pressure 
\cite{pressure-2014,pressure-2015,pressure-2015b}.  
%Particles become de-adsorbed only when a direction of 
%a new velocity points away from a wall.  This effect could be incorporated into Eq. (\ref{eq:FP}) 
%by introducing sinks and sources at the location of the walls, but things then start to become 
%complicated.  
This could be handled by introducing sinks and sources in the Fokker-Planck equation, 
and such approach was used for the case of active Brownian particles \cite{NJP-Lee-2013,JSTAT-Wagner-2017}, 
an alternative model of active particles wherein orientation of a swimming velocity changes by diffusion.  
The resulting mathematics, however, is rather complex.  
To our knowledge, no similar study has been done for RTP particles between two walls for an 
arbitrary dimension.  

%In this work, we analyze RTP particles between two walls by reinterpreting
%the system as the splitting probability problem, which allows us to analyze it in a different and more
%convenient framework.   

%A different approach is to extend the delta distribution in Eq. (\ref{eq:pv}) (for the case $d=1$) to $n$ 
%number of discrete velocities, until reaching the limit $n\to \infty$ to find the solution for a continuous 
%distribution $p_v$.   The problem with this approach is that mathematical complexity increases very 
%quickly.  Already analytical expressions for the four-state model become rather large \cite{Frydel22b}, 
%and going beyond four-states does not seem like a good idea.  The seemingly simple problem of RTP 
%particles between two walls, therefore, remains an unresolved, 
%
%In this work we propose an alternative way of looking at the stationary system of RTP particles between 
%two walls.  This reinterpretation permits us to formulate the system 
%not through the Fokker-Planck equation but as an integral equation involving a well 
%defined Green's like function.       

\section{redefining the microscopic process}
%{from Langevin dynamics to jump-process algorithm}

The usual way to simulate RTP motion is by integrating the Langevin equation which, 
in the absence of an external potential and using the Euler method, amounts to the following 
formula
\be
x(t+\Delta t) =   x(t) + v \Delta t,
\label{eq:langevin-0}
\ee
where the velocity $v$ is drawn from the distribution $p_v$, given in Eq. (\ref{eq:pv}), at time intervals $t_p$ 
drawn from $p_t$ in Eq. (\ref{eq:pt}).

If the objective is to study dynamics, then the integration of a Langevin equation is 
the only option to simulate the system.  
%One can change a numerical integration method to a more 
%sophisticated one involving higher order terms, but the basic idea is the same.  
But if the interest is in a stationary state, then we might have some flexibility in how to sample 
%forgo the explicit evolution in time based on the Langevin equation and 
different configurations.  
%For example, for a system in equilibrium we have a choice between Monte 
%Carlo sampling and molecular dynamics.  
%%Why not do something similar for a non-equilibrium stationary system.  
%This sounds like a good general idea, but the question is, how to find such an algorithm.  

In this section we propose an alternative microscopic process that yields the same stationary 
distribution as that generated by the RTP motion.  
%We propose such an algorithm.  
Rather than tracing the position of a particle at each moment, we 
consider an algorithm that generates configurations at the "tumble" stage and ignores 
configurations generated during the "run" stage.  As a result, rather than moving continuously, 
a particle appears to be jumping from one place to another.  
%This results in a particle that no longer 
%moves continuously through a medium but is jumping.  

If at a given "tumble" stage, occurring at time $t$, a particle is located at $x(t)$, then its position at 
the subsequent "tumble" stage is 
\be
x(t+t_p) =  x(t)    +    v t_p.  
\label{eq:xtp}
\ee
A particle makes a jump of length $\Delta x = vt_p$, where $\Delta x$ is a product of 
two random variables $v$ and $t_p$.  Note that the sequential jumps do not correspond to 
the passage of time of the original RTP system.  $t_p$ in this algorithm is just a random parameter 
used to generate a next jump. 

To get rid of any appearance of time, we represent Eq. (\ref{eq:xtp}) as 
\be
x_{n+1} =  x_n    +   v t_p,
\label{eq:scheme-0}
\ee
where $x_{n}$ is a position of a particle at step $n$.  We refer to the algorithm based on 
Eq. (\ref{eq:scheme-0}) as the "jump-process" algorithm.  
%Note that the system we study 
%continues to be the RTP model.  We simply modify the way in which configurations are sampled.  
%And since the configuration sampling does not obey any internal clock, the algorithm no longer 
%qualifies as representing molecular dynamics.    

\begin{figure}[hhhhh] 
 \begin{center}
 \begin{tabular}{rrrr}
 \includegraphics[height=0.19\textwidth,width=0.21\textwidth]{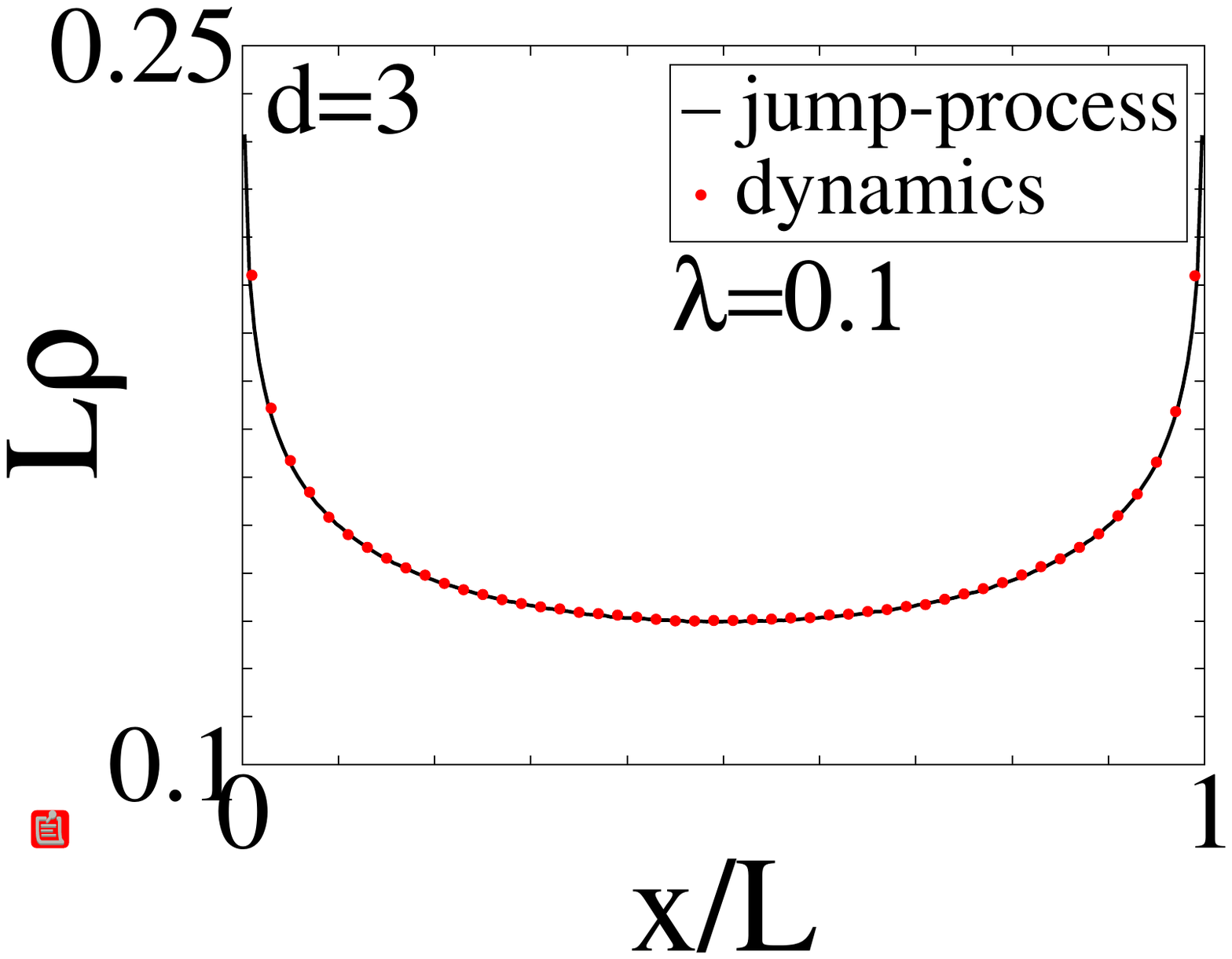} &
 \includegraphics[height=0.19\textwidth,width=0.21\textwidth]{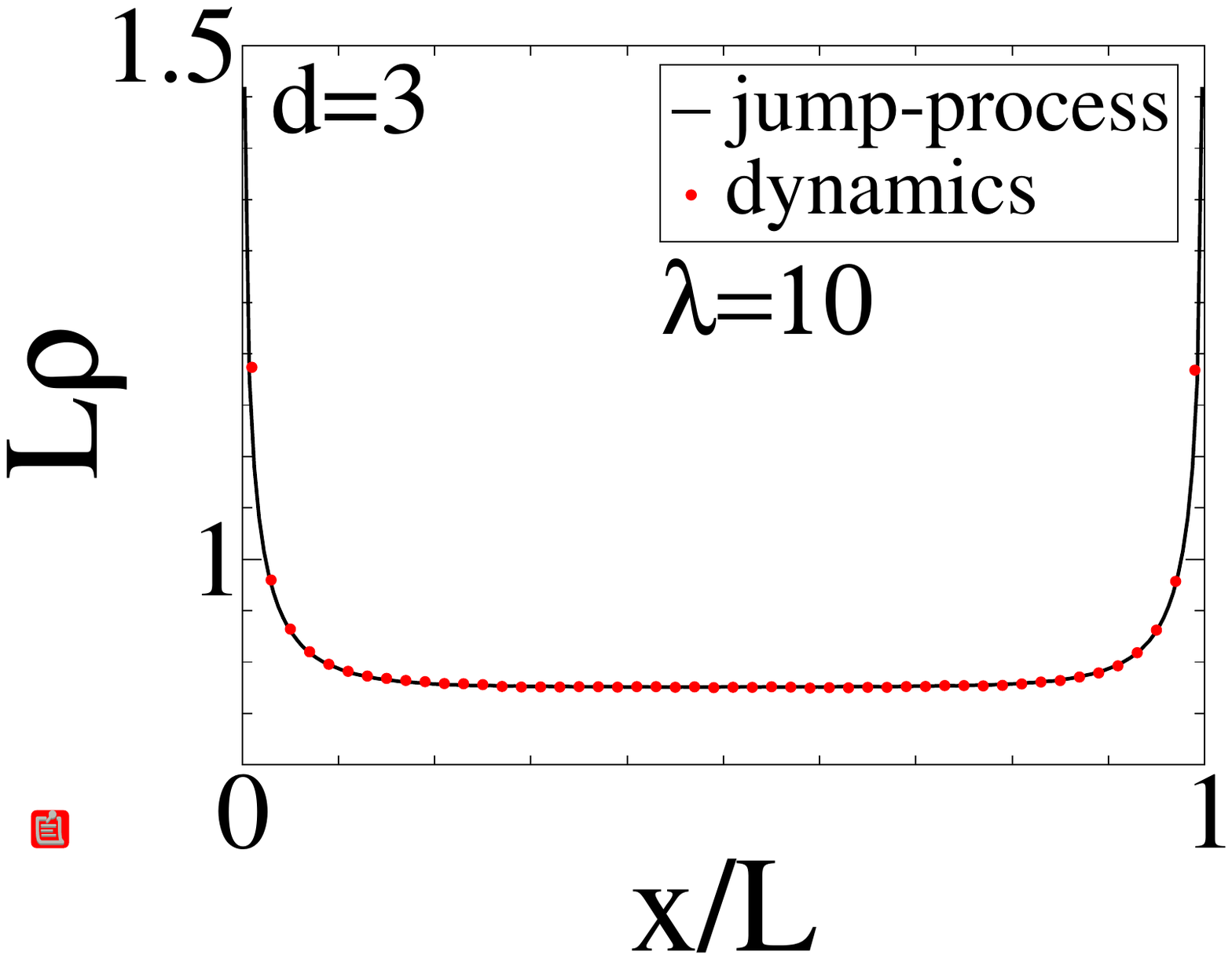} \\
 \includegraphics[height=0.19\textwidth,width=0.21\textwidth]{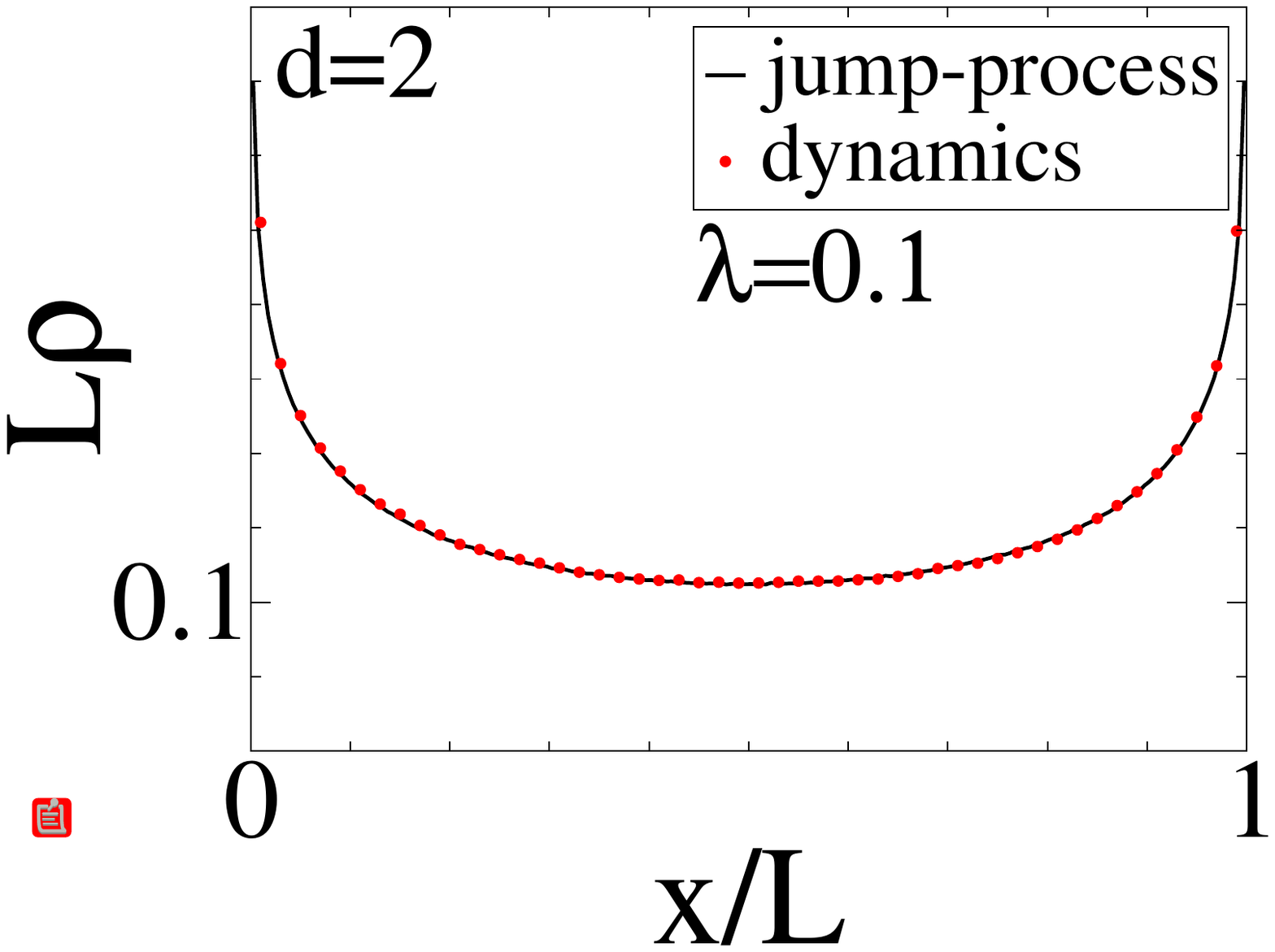} &
 \includegraphics[height=0.19\textwidth,width=0.21\textwidth]{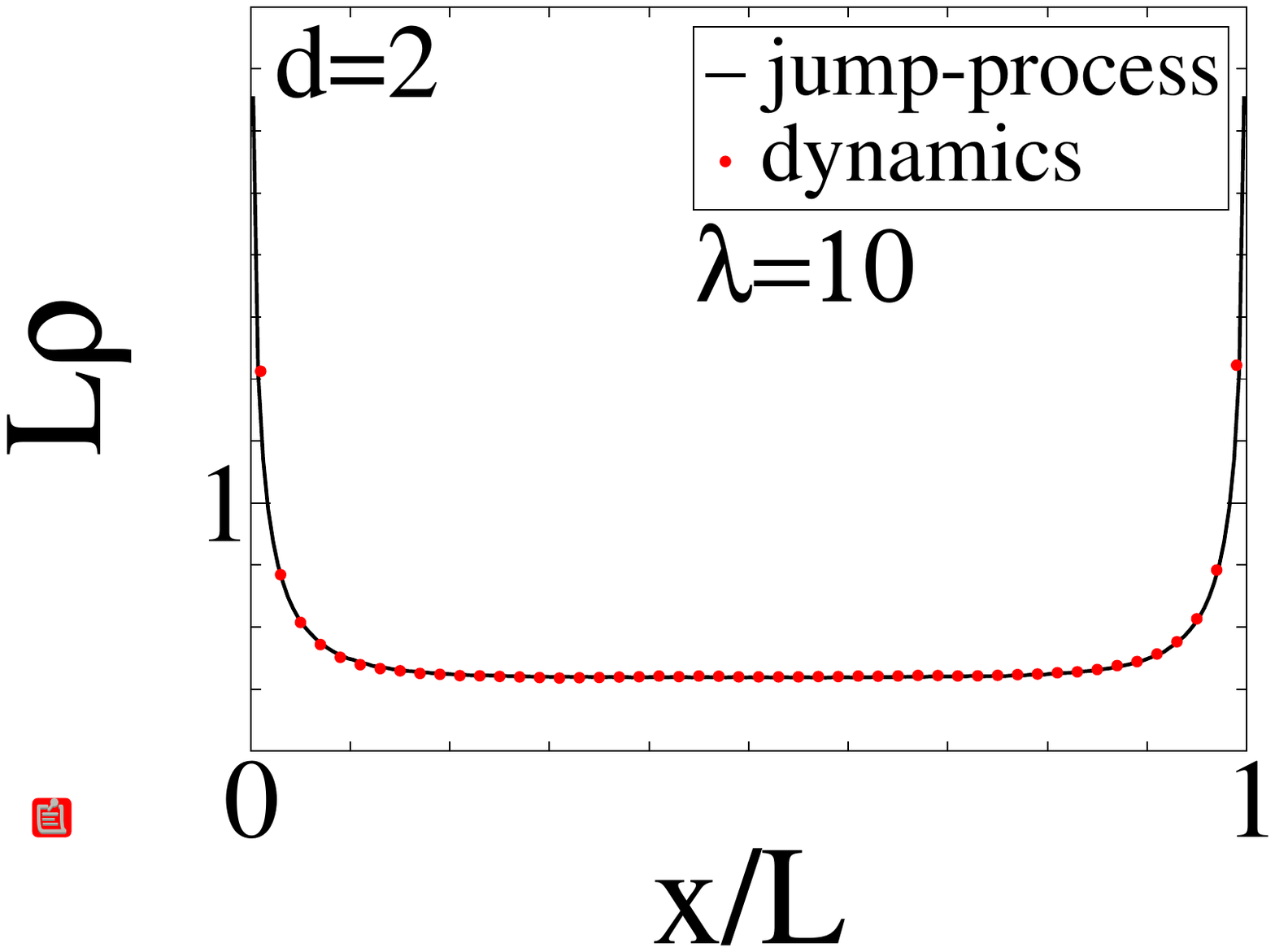} \\
\includegraphics[height=0.19\textwidth,width=0.21\textwidth]{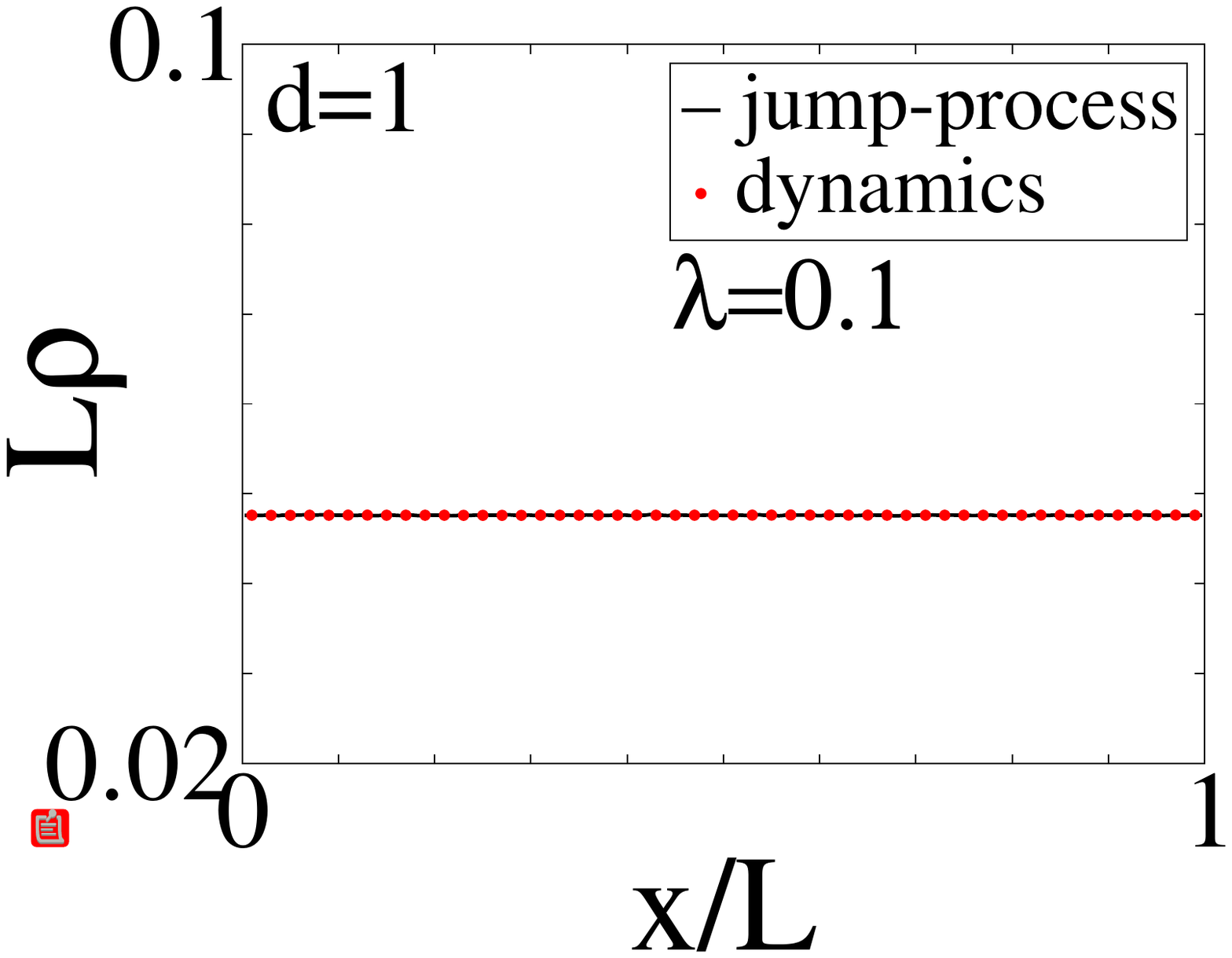}  &
 \includegraphics[height=0.19\textwidth,width=0.21\textwidth]{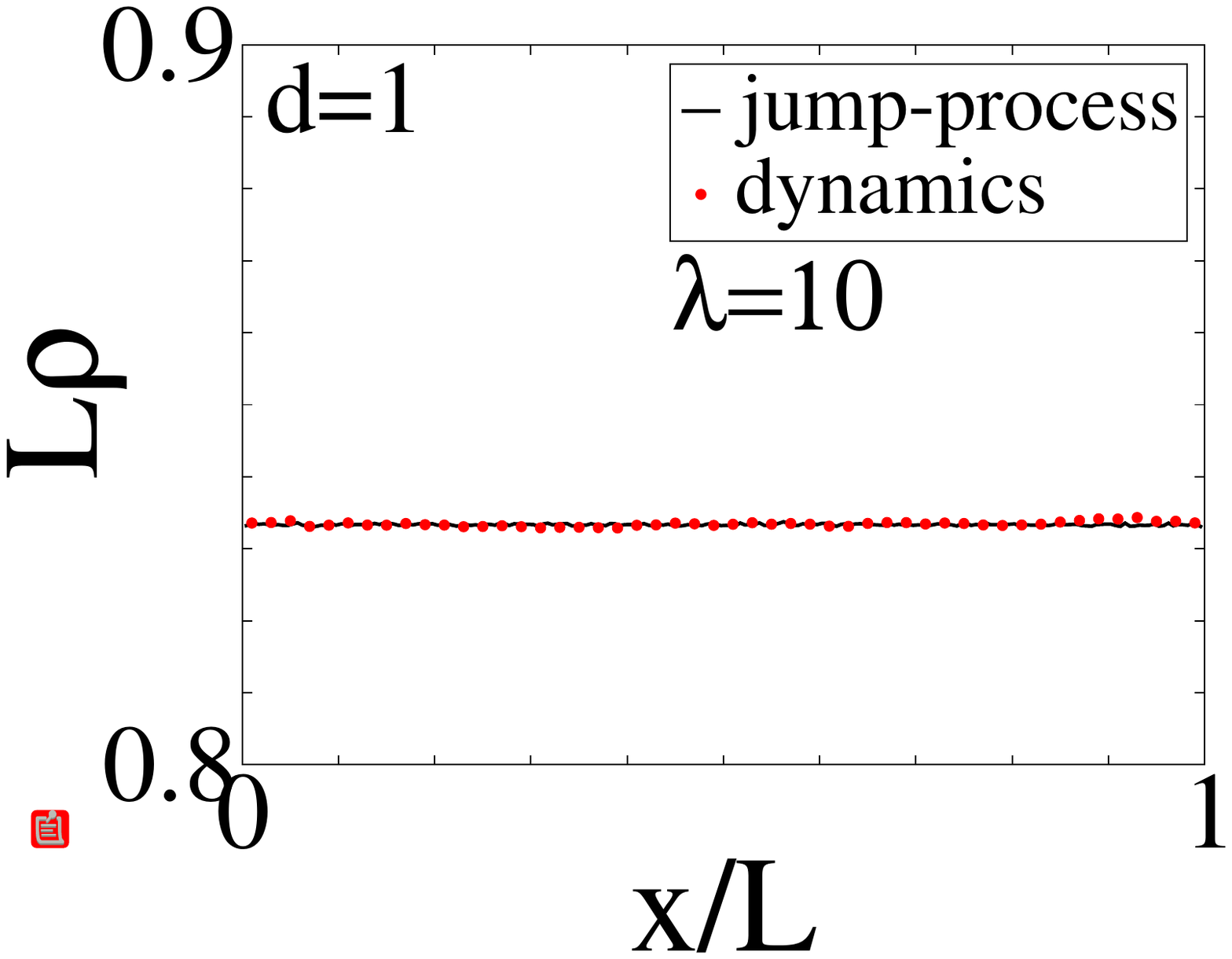} 
\end{tabular}
 \end{center} 
\caption{Stationary (marginal) distributions of RTP particles confined between two walls at $x=0$ and $x=L$, 
for different dimension $d$ and $\lambda=\frac{L}{\tau v_0}$.    
Each distribution is generated by two sampling algorithms.  Distributions $\rho$ represent
non-adsorbed particles only.}
\label{fig:rho-wall} 
\end{figure}

In Fig. (\ref{fig:rho-wall}) we compare stationary distributions generated by the Langevin dynamics of 
Eq. (\ref{eq:langevin-0}) with stationary distributions generated using the "jump-process" algorithm of 
Eq. (\ref{eq:scheme-0}) for different values of $\lambda$ defined as
$$
\lambda = \frac{L}{\tau v_0},
$$
and for different dimension $d$.  
%%%%%%%%%%%%%%%%%%%%%%
\graphicspath{{figures/}}
%%%%%%%%%%%%%%%%%%%%%%%
Distributions (corresponding to the same physical parameters) generated by different sampling 
methods are indistinguishable apart from statistical error, confirming the correctness of the 
"jump-process" algorithm.  Apparently, by eliminating the "run" stage of the motion, we do not 
%lose information, or alter in any way the 
alter statistical weight of different configurations.  

In both algorithms, the interaction with the walls is implemented as follows.  Every time a particle crosses 
one of the walls at $x=0$ or $x=L$, it is moved to the location of the wall it just crossed.  This particle is 
considered to be adsorbed.  If an adsorbed particle crosses a wall at the next step, it is brought 
back to the location of a wall, thus, it remains adsorbed.  
The procedure is repeated until a particle moves into the region $x\in (0,L)$, 
in which case it is considered as de-adsorbed.

Since in a stationary state there is always a fraction of particles that is adsorbed, distributions in Fig. (\ref{fig:rho-wall}) 
do not integrate to one but $\int_0^Ldx\, \rho(x) = 1- f_A$ where $f_A$ is the fraction of adsorbed 
particles, and the distribution of adsorbed particles is represented by two delta functions at the location 
of the walls, 
\be
\rho_A = \frac{f_A}{2} \delta(x)   +   \frac{f_A}{2} \delta(x-L).  
\label{eq:rho-ads}
\ee
It is helpful to think of particles as those that are adsorbed and those that are free, as 
%The phenomenon of adsorption is a manifestation of common property of active particles
%to accumulate near trap borders.  
%This separation of particles into those that are adsorbed
%and those that are free.  
each type exhibits different behavior, for example, they dissipate
heat differently.  In addition, adsorbed particles are responsible for exerting pressure 
by pushing against the walls.

Fig. (\ref{fig:rho-wall}) illustrates how stationary distributions vary with dimension.  For $d=1$, 
all distributions are uniform and the only quantity that changes with $\lambda$ is $f_A$ and the system
for this dimension 
is fully analytically tractable \cite{Malakar-2018,PRE-Razin-2020}.  The structure of $\rho$ for higher
dimensions is more complex.  One distinct feature is the emergence of divergences at the location 
of the walls.  
%To our knowledge, no exact analytical results are available for $d>1$. 

\section{analysis of jump-process algorithm}

Given a length of a single jump (that depends on two independent random variables $t_p$ and $v$) 
\be
\Delta x = v t_p, 
\label{eq:jump-length}
\ee
the probability distribution of jump lengths is defined as 
\be
G(\Delta x) = \int_0^{\infty} dt_p \, p_t (t_p) \int_{-v_0}^{v_0} dv \, p_v(v) \delta\left( \Delta x -  v t_p \right). 
\label{eq:G-00}
\ee
The integral in Eq. (\ref{eq:G-00}) is easily evaluated due to the presence of a delta function which 
eliminates one of the integrals. 
The evaluated expression depends on a system dimension via $p_v$ in Eq. (\ref{eq:pv}).  This 
leads to three different probability distributions given by 
\be
G(\Delta x) 
=  \frac{1}{2} \frac{ 1 }{\tau v_0} 
\begin{cases}
    e^{-|\Delta x| / \tau v_0}, & \text{for $d=1$},\\
    \frac{2}{\pi}  \text{K}_0\big(|\Delta x| / \tau v_0 \big), & \text{for $d=2$}, \\
     \Gamma\big( 0, |\Delta x| / \tau v_0 \big)  , & \text{for $d=3$}, \\
  \end{cases}
\label{eq:G-0}
\ee
where $\text{K}_n(x)$ is the modified Bessel function of the second kind and $\Gamma(n,x)$ is the 
incomplete gamma function.  The distributions for $d=2$ and $d=3$ diverges at $\Delta x = 0$.  
In both cases the singularity is logarithmic,   
\ba
&& \text{K}_0 \sim -\ln \frac{|\Delta x|}{2\tau v_0} - \gamma  \nonumber\\
&& G_0 \sim   -\ln \frac{|\Delta x|}{\tau v_0} - \gamma,  
\label{eq:sing}
\ea
where $\gamma$ is the Euler's constant.  
Divergence for $d>1$ can be explained by the fact that $p_v$ in those cases does not vanish at 
$v=0$.  This increases the probability for a particle not to move at the next jump.  
A logarithmic divergence is integrable and the second 
moment of $G(\Delta x)$ exists and is given by 
\be
\langle \Delta x^2\rangle = \frac{2\tau^2 v_0^2}{d}.  
\label{eq:SD}
\ee
According to this result, the variance decreases with increasing dimension.

%Since the 
%jump length is a product of two random variables, see Eq. (\ref{eq:jump-length}), and since for 
%the case $d>1$ both $t_p$ and $v$ can be zero, this considerably increases the probability for 
%the jump length $\Delta x\approx 0$, leading to the logarithmic divergence.  

\section{Stationary distribution from $G(\Delta x)$}

%Having derived exact expressions for the jump probability distributions in Eq. (\ref{eq:G-0}), 
%we can now calculate a stationary marginal distribution that results from this process
%for a particle confined to the region $x\in(0,L)$. 
Once we have the exact expression for the probability distribution of jumps, we can calculate 
a stationary distribution of non-adsorbed particles 
%can be derived expressed as 
from the following integral equation:    
\ba
\rho(x) &=& \int_0^{L} dx' \, \rho(x') G(x'-x)   \nonumber\\
&+&  \frac{f_{A}}{2} G(x) + \frac{f_{A}}{2}  G(x-L). %\nonumber\\
\label{eq:rho-0}
\ea
The last two terms are contributions of recently de-adsorbed particles, where the factor $f_A/2$ is 
the fraction of particles adsorbed onto a single wall.  Because $\rho$ in Eq. (\ref{eq:rho-0}) does not 
include adsorbed particles, it is normalized as 
\be
%f_A  = 1 - \int_0^L dx\, \rho
\int_0^L dx\, \rho = 1 - f_A.  
\label{eq:fA}
\ee

\begin{figure}[hhhh] 
 \begin{center}
 \begin{tabular}{rrrr}
 \includegraphics[height=0.19\textwidth,width=0.21\textwidth]{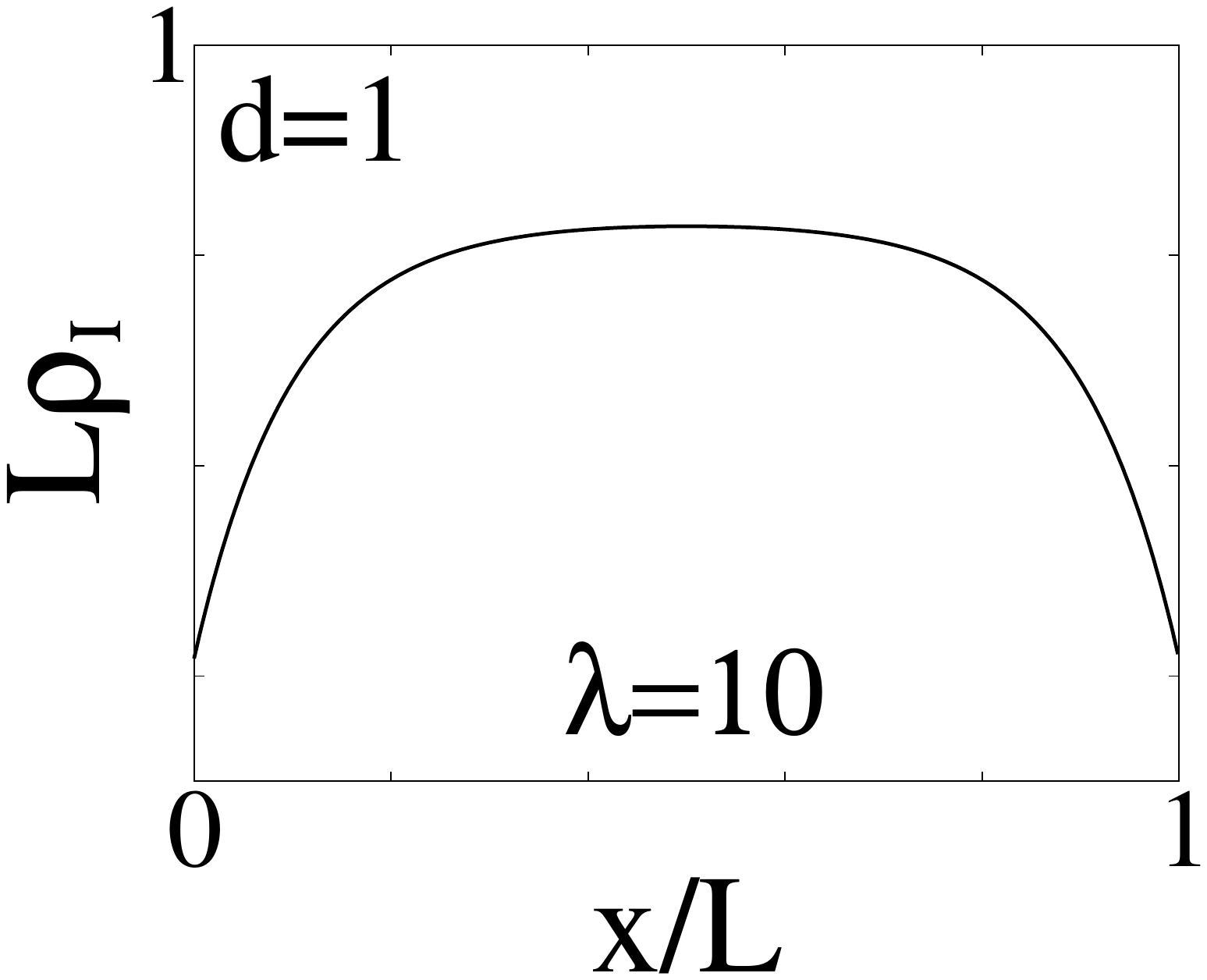} &
 \includegraphics[height=0.19\textwidth,width=0.21\textwidth]{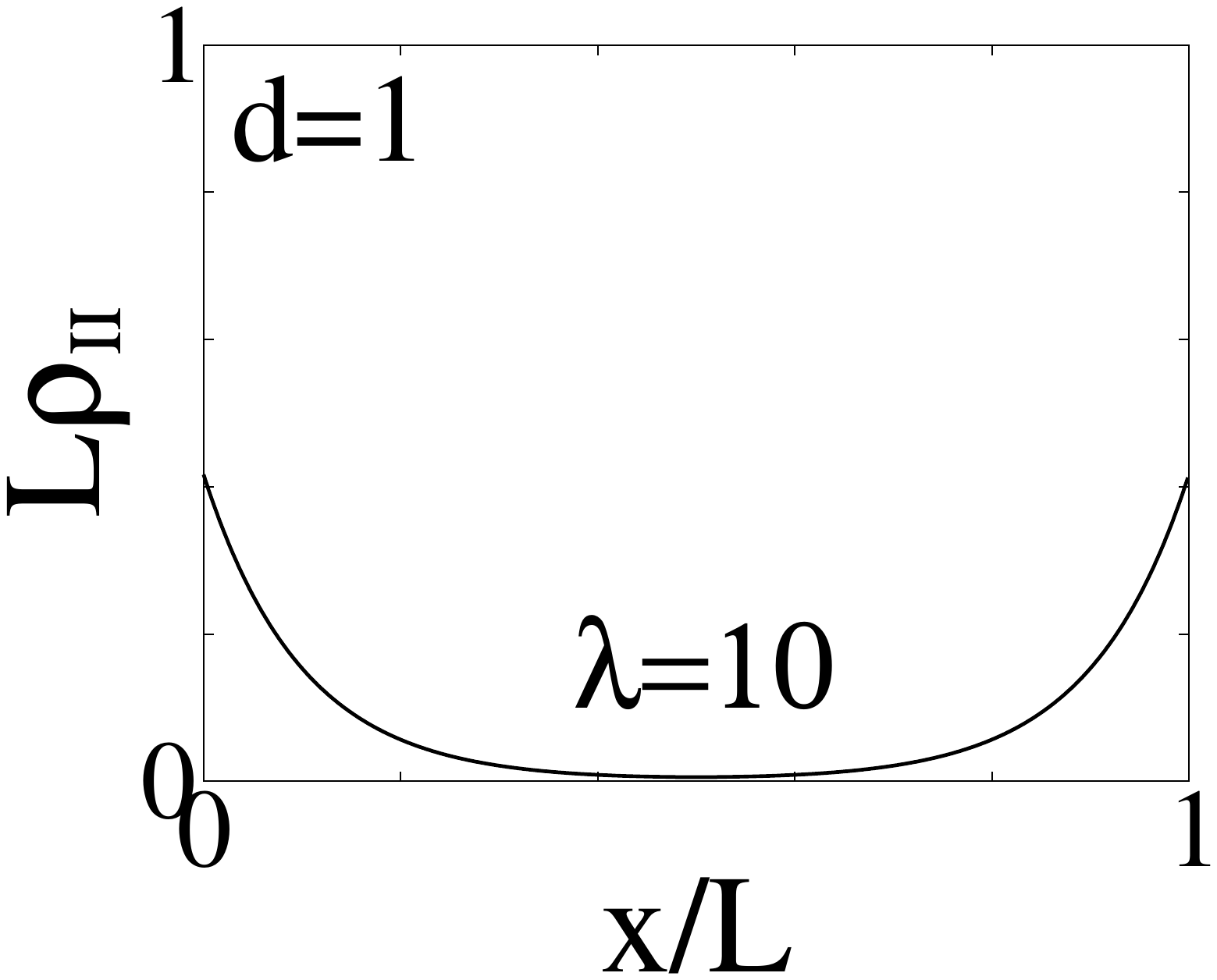} \\
 \includegraphics[height=0.19\textwidth,width=0.21\textwidth]{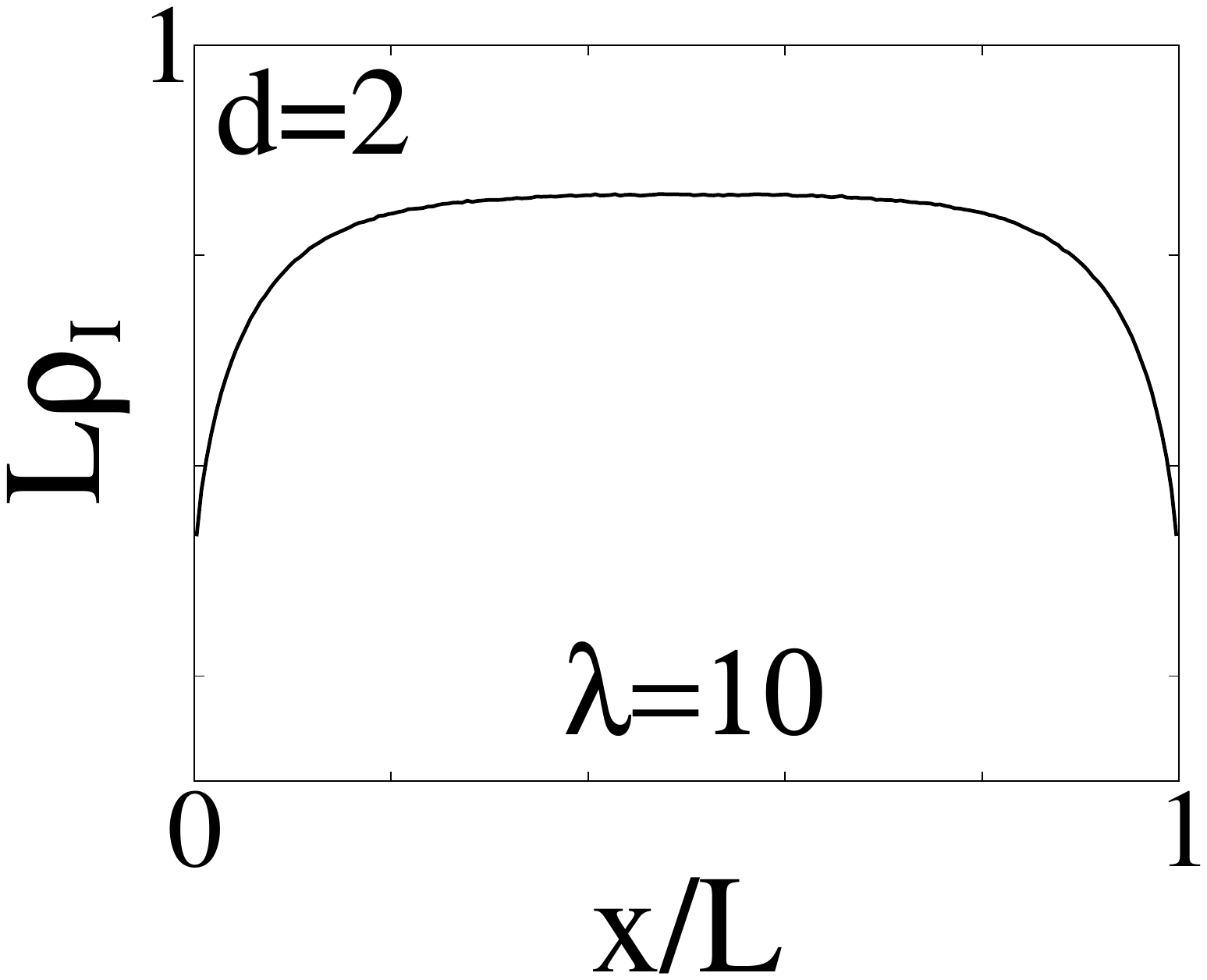} &
 \includegraphics[height=0.19\textwidth,width=0.21\textwidth]{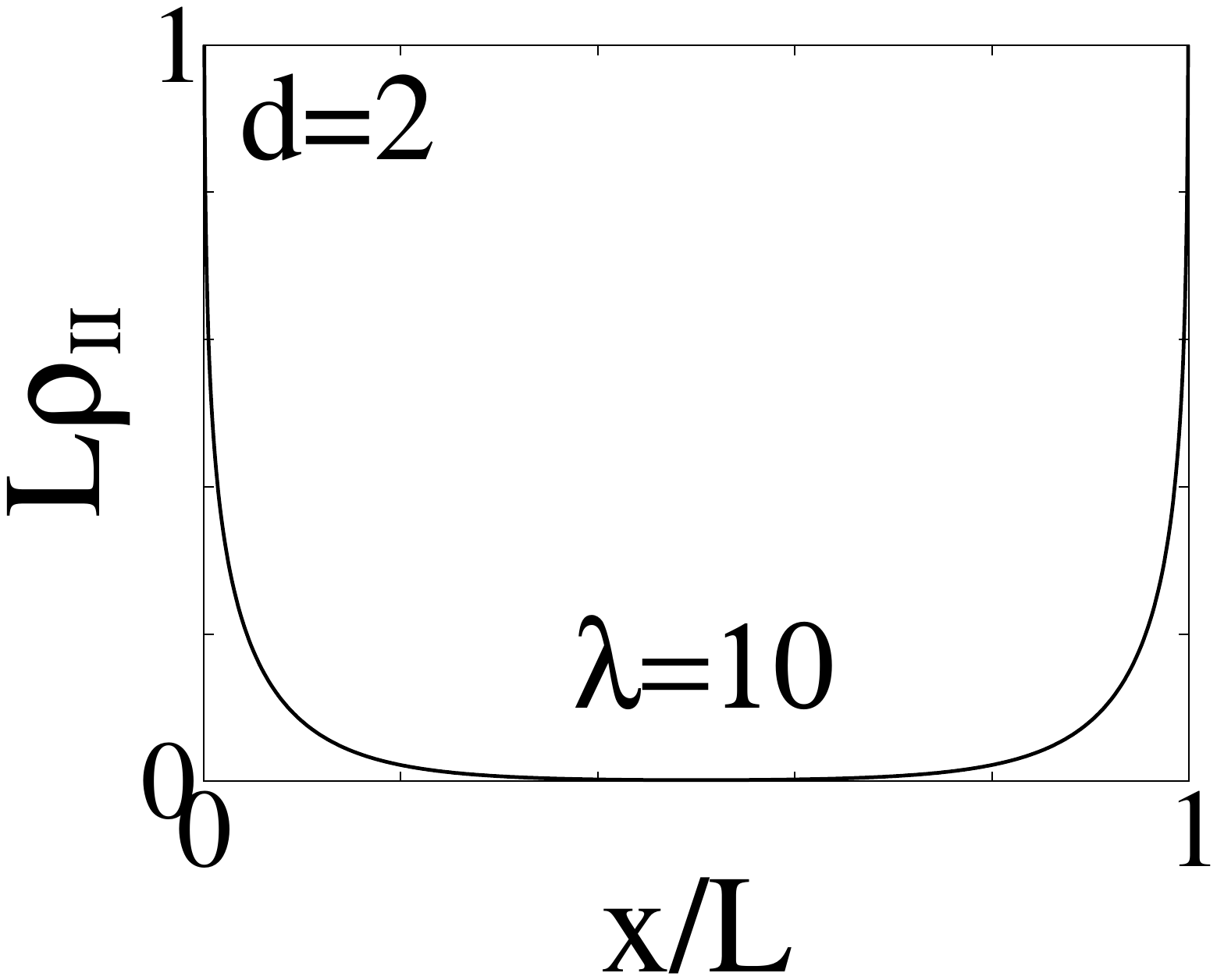} \\
 \includegraphics[height=0.19\textwidth,width=0.21\textwidth]{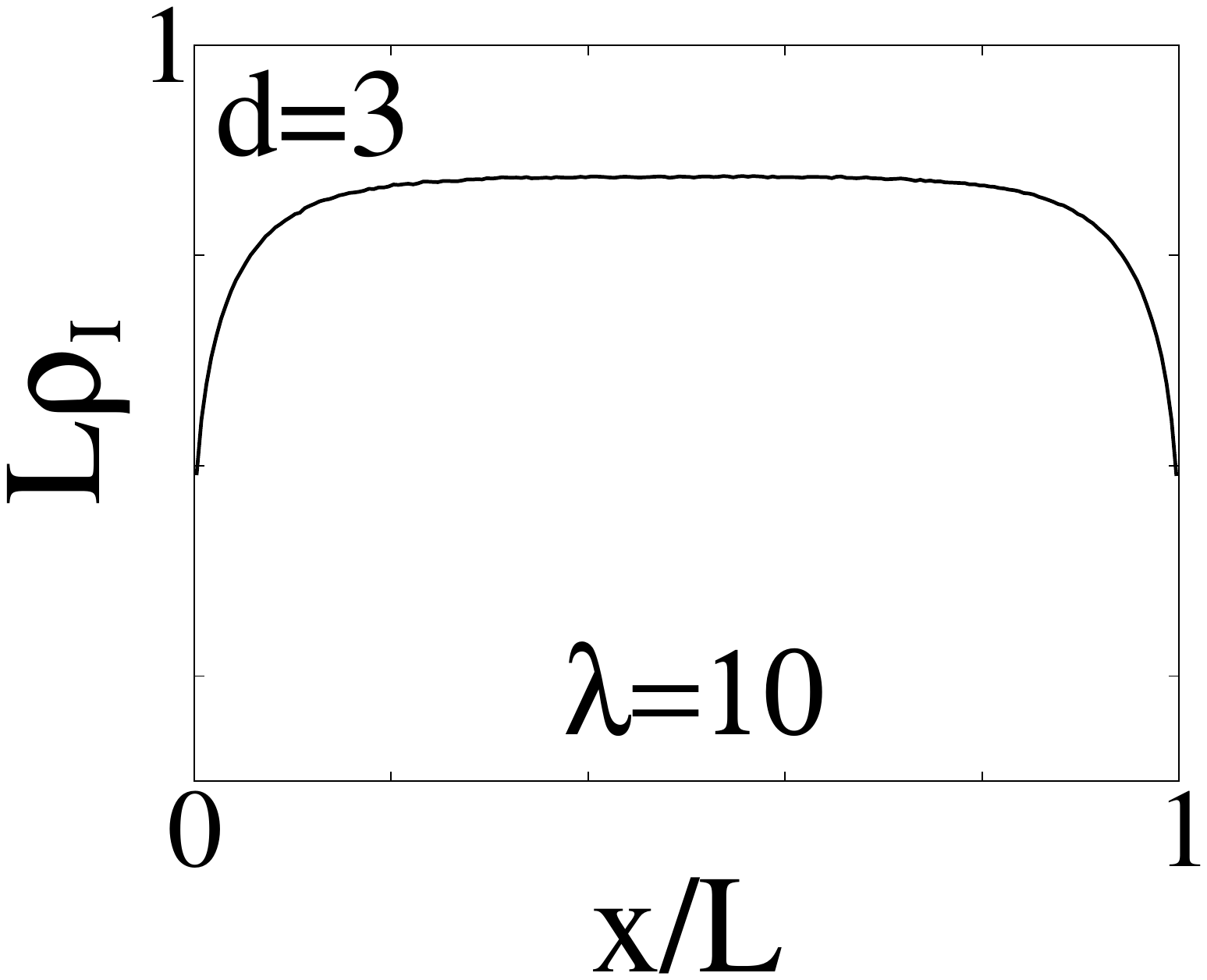} &
 \includegraphics[height=0.19\textwidth,width=0.21\textwidth]{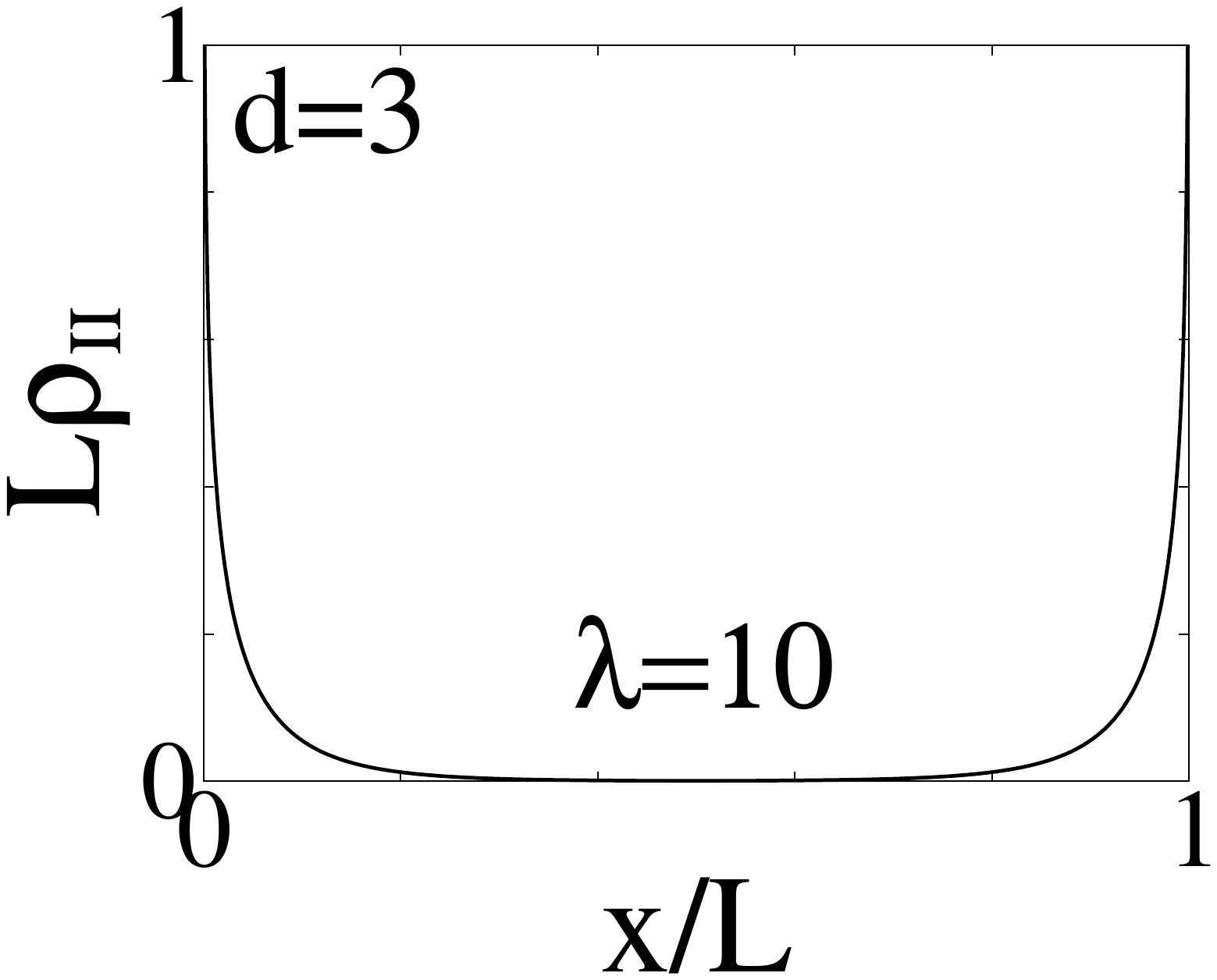} \\
\end{tabular}
 \end{center} 
\caption{Contributions to a stationary distributions $\rho$.   $\rho_I$ corresponds to the first line and $\rho_{II}$ corresponds 
to the second line  in Eq. (\ref{eq:rho-0}).   The dimensionless parameter $\lambda$ is defined 
as $\lambda = \frac{L}{\tau v_0} $. } 
%$\tau=0.1$, $L=1=v_0=1$ } 
\label{fig:rho-wall-all} 
\end{figure}

Even without solving Eq. (\ref{eq:rho-0}), we can already extract useful results from it.  For example,
a divergence at each wall for $d>1$, seen in Fig. (\ref{fig:rho-wall}), can be linked to  
distributions $G$ in the second line of Eq. (\ref{eq:rho-0}) and representing the distribution of 
recently re-adsorbed particles.  
To confirm that the diverge comes only from those terms, 
in Fig. (\ref{fig:rho-wall-all}) we plot the contributions 
%different contributions of Eq. (\ref{eq:rho-0}), 
$\rho_I  = \int_0^{L} dx_0 \, \rho(x_0) G(x-x_0)$ and 
$\rho_{II} = \frac{f_{A}}{2} G(x) + \frac{f_{A}}{2}  G(x-L)$, such that $\rho = \rho_I + \rho_{II}$. 
%%%%%%%%%%%%%%%%%%%%%%
\graphicspath{{figures/}}
%%%%%%%%%%%%%%%%%%%%%%%
To plot $\rho_{II}$, we need to calculate $f_A$, which is obtained
from a "jump-process" simulation.  
To calculate $\rho_I$, we use a "jump-process" algorithm in which we ignore 
particle positions of recently de-adsorbed particles, since these 
configurations contribute to $\rho_{II}$.

Regardless of a system dimension, all distributions $\rho_I$ exhibit depletion near the two walls.  
The depletion arises since walls act like an adsorbing boundary.  Since $\rho_{I}$ does not diverge anywhere
in the region $x\in[0,L]$, we conclude that a divergence comes from 
$G$ at the location of the two walls, and whose exact expression is given in Eq. (\ref{eq:G-0}).

For the case of an exponential distribution $G$, corresponding to $d=1$, Eq. (\ref{eq:rho-0})
can be solved exactly.  A stationary distribution in this case is uniform, as seen in Fig. (\ref{fig:rho-wall}), 
given by $\rho=(1-f_A)/L$.  Inserting this into Eq. (\ref{eq:rho-0}) yields a formula for $f_A$ given by 
\be
f_A = \frac{2}{2+\lambda}.  
\label{eq:fA-1d}
\ee

\section{Analogy with the splitting probability problem}

%In this section we discuss the analogy or the re-interpret the RTP system as 

By redefining the RTP motion as a jump process, we inadvertently link the RTP system of particles 
confined between two walls 
to the splitting probability problem \cite{PRL-Klinger-2022}.  The splitting probability $\pi(x)$ 
is the probability that a particle starting at $x\in [0,L]$ reaches a wall at $x=L$ without first being 
adsorbed onto a wall at $x=0$.  
The splitting probability $\pi(x)$ satisfies the following integral equation 
%has a similar structure to Eq. (\ref{eq:rho-0}) and is given by 
\cite{PRL-Klinger-2022}
\be
\pi(x)   =    \int_{0}^{L} dx'\, \pi(x') G(x'-x)   +   \int_{L}^{\infty} dx'\, G(x'-x).  
\label{eq:pi}
\ee
The second term in the above equation is the probability of reaching the wall at $x=L$ in a single jump, 
and the first term is the probability of reaching the same wall in multiple jumps.  

To make the analogy between the splitting probability problem and the stationary RTP system  
more obvious, we differentiate Eq. (\ref{eq:pi}) with respect to $x$.  This leads to 
\ba
\frac{d \pi }{dx}    &=&    \int_{0}^{L} dx'\, G(x'-x)  \frac{d \pi }{dx'}   \nonumber\\ 
&+&   \pi(0) G(x)  +  \Big[1 -  \pi(L) \Big] G(x-L).
\ea
Comparing the above equation with Eq. (\ref{eq:rho-0}), we establish the following identities:  
\be
\rho = \frac{d \pi}{dx},
\label{eq:rho-pi}
\ee
indicating that $\rho(x)$ is the derivate of $\pi(x)$, and the boundary conditions of the probability 
$\pi(x)$ are found to be related to the fraction of adsorbed particles, 
\ba
&& \pi(0) = \frac{f_A}{2}  \nonumber\\ 
&& \pi(L) = 1-\frac{f_A}{2}.  
\label{eq:fA-pi}
\ea
One advantage of Eq. (\ref{eq:pi}) is that $\pi(x)$ does not diverge at the walls.  The equality 
$\pi(0)=f_A/2$ is especially interesting.  It tells us that the fraction of adsorbed particles at a 
wall is the same as the probability of reaching that wall by a particle that initially is adsorbed 
at another wall, without being re-adsorbed during its random walk.

The reinterpretation of the RTP system as the splitting probability problem, 
%transformation of formulation from the stationary Fokker-Planck equation to 
formulated as the integral equation in Eq. (\ref{eq:pi}), together with the relations in 
Eq. (\ref{eq:rho-pi}) and Eq. (\ref{eq:fA-pi}), is the main result of this work.  The important point is 
that Eq. (\ref{eq:pi}) is not derived from the Fokker-Planck equation but is obtained by changing the 
microscopic process iteslf.  
%The macroscopic theory is subsequently derived based on this process. 
%from continuous dynamics to discontinuous jumps, and then building a  
%macroscopic theory based on that process.  

%The determination of the microscopic process that is compatible with the original RTP dynamics
%was not exactly derived but was more of a result of a trial and error procedure.  We tried it 
%and it worked.  This shows that for a stationary system, configurations generated during the 
%"run" stage of the RTP process can be discarded.  Configurations can be sampled by 
%the "tumble" stage alone.  

Certain properties of $\pi(x)$ can be established from symmetry consideratons.   
For example, for a particle starting from the middle of the interval, the splitting distribution is $\pi(L/2)  =  {1}/{2}$, since 
the likelihood of reaching either wall is the same.  Another relation 
from symmetry is $\pi(L/2-x)  +   \pi(L/2+x)  =  1$.  Combining the two previous relations we can establish 
$\frac{1}{L} \int_0^{L} dx\, \pi(x)  =  \frac{1}{2}$, which means that considering all starting points
on $(0,L)$, there is on average fifty percent chance that a particle will reach a wall at $x=L$.

The integral in Eq. (\ref{eq:pi}) can be expanded 
%as an infinite series of integrals of increasing order 
%involving $G$.  Such expression can be obtained 
by repetitive elimination of $\pi(x)$ on the 
right-hand-side with the initial terms of expansion given by 
%\ba
%\pi(x)   &=&    \int_{L}^{\infty} dx'\, G(x'-x)          \nonumber\\
%&+&  \int_{0}^{L} dx'\,  G(x'-x)  \int_{L}^{\infty} dx''\, G(x''-x')           \nonumber\\
%&+&  \int_{0}^{L} dx'\,   G(x'-x)  \int_{0}^{L} dx''\, G(x''-x')   \int_{L}^{\infty} dx'''\, G(x'''-x'')     \nonumber\\
%&+&  \dots
%\ea
\ba
\pi(x)   &=&    \int_{L}^{\infty} dx'\, G(x-x')          \nonumber\\
&+&  \int_{0}^{L} dx'\,   G(x-x')  \int_{L}^{\infty} dx''\, G(x'-x'')           \nonumber\\
&+&  \int_{0}^{L} dx'\,   G(x-x')  \int_{0}^{L} dx''\, G(x'-x'')   \int_{L}^{\infty} dx'''\, G(x''-x''')     \nonumber\\
&+&  \dots
\ea
This allows us to express $\pi(0)$ as $\pi(0) = \sum_{n=1}^{\infty} p_n$ where $p_n$ is the probability to reach 
%represent $\pi(0)$ as a sum of probabilities that 
the wall at $x=L$ 
%is reached 
in $n$ jumps without being re-adsorbed.  
%leading to the formula $\pi(0) = \sum_{n=1}^{\infty} p_n$,  
Explicit expressions for $p_n$ for initial terms is given below 
%\ba
%&&p_1 = \int_{L}^{\infty} dx\, G(x)  \nonumber\\
%%=  \frac{1}{2} - \int_{0}^{L} dx\, G(x,0) 
%%&&p_2 = \int_0^L dx_0\, G(x_0) \int_{L}^{\infty} dx\, G(x_0-x)  \nonumber\\
%&&p_2 = \int_0^L dx'\, G(x') \int_{L}^{\infty} dx\, G(x'-x)  \nonumber\\
%%&&p_3 = \int_0^L dx_0\, G(x_0)  \int_0^L dx_1 \, G(x_0-x_1) \int_{L}^{\infty} dx\, G(x_1-x)  \nonumber\\
%&&p_3 = \int_0^L dx''\, G(x'')  \int_0^L dx' \, G(x''-x') \int_{L}^{\infty} dx\, G(x'-x).  \nonumber\\
%%&&    \dots
%\label{eq:pn}
%\ea
\ba
&&p_1 = \int_{L}^{\infty} dx'\, G(x')  \nonumber\\
&&p_2 = \int_0^L dx'\, G(x') \int_{L}^{\infty} dx''\, G(x'-x'')  \nonumber\\
&&p_3 = \int_0^L dx'\, G(x')  \int_0^L dx'' \, G(x'-x'') \int_{L}^{\infty} dx'''\, G(x''-x''').  \nonumber\\
\label{eq:pn}
\ea

\subsection{the case $d=1$}

%The case $d=1$ is analytically tractable and the results can offer useful insights.  
Given that for $d=1$, $\rho$ is uniform and given by $\rho = (1-f_A)/L$, 
%the relation in Eq. (\ref{eq:rho-pi}) implies that 
$\pi(x)$ can be obtained from Eq. (\ref{eq:rho-pi}) 
%must be linear in $x$, and to satisfy 
and the boundary conditions in Eq. (\ref{eq:fA-pi}), leading to 
\be
\pi(x)   =    \frac{1}{2}  +   \left( 1-f_A \right) \left( \frac{x}{L}  -  \frac{1}{2} \right).  
\label{eq:pi-1d}
\ee
%Note that apart from the boundary conditions the above relation satisfies $\pi(L/2)=1/2$.  
Inserting the above expression into Eq. (\ref{eq:pi}) yields 
\be
\frac{f_A}{2} = \frac{1}{2+\lambda}, 
\label{eq:fA-1d-R}
\ee
which recovers the result in Eq. (\ref{eq:fA-1d}) and agrees with an alternative derivation based on 
%confirms the same result 
%which can also be 
%derived by 
solving a stationary 
Fokker-Planck equation \cite{PRE-Razin-2020,PRE-Fyrdel-2022}.  
%This proves the accuracy of the 
%"jump-process" algorithm for the system in $d=1$, not just through the comparison of simulation 
%results as done in Fig. (\ref{fig:rho-wall}).  

Since the system in $d=1$ is analytically tractable, we can obtain other exact results.  
For example, for a particle starting at one of the walls, an expression for the average number of steps 
needed to reach the opposite wall, defined as 
\be
\langle n\rangle =  \frac{2}{f_A} \sum_{n=1}^{\infty} n\,p_n,
\ee
can be obtained by generating initial terms of the expansion 
$\langle n\rangle = c_0 + c_1 \lambda + \dots$ and then extracting from those terms a general 
sequence function.  
For the exponential distribution $G$ corresponding to $d=1$ 
%this is not a difficult task.  All the 
%probabilities $p_n$ in this case are a function of a single parameter $\lambda$ with 
the expansion of $p_n$ starts from 
\be
p_n(\lambda) = \frac{\lambda^{n-1}}{2^n}  + O(\lambda^{n}).  
\label{eq:pn-expan}
\ee
From this it follows that the $\lambda$-expansion of $\langle n^k \rangle$ 
can be represented as 
\ba
\frac{f_A}{2} \langle n^k \rangle &=&  p_1(0)  \nonumber\\
&+&  \lambda \lim_{\lambda\to 0} \left[ \frac{d p_1}{d\lambda} + 2^k \frac{dp_2}{d\lambda}  \right] \nonumber\\
&+&  \frac{\lambda^2}{2!} \lim_{\lambda\to 0} \left[ \frac{d^2 p_1}{d\lambda^2} + 2^k \frac{d^2p_2}{d\lambda^2}  + 3^k \frac{d^2p_3}{d\lambda^2}   \right] \nonumber\\
&+&  \frac{\lambda^3}{3!} \lim_{\lambda\to 0} \left[ \frac{d^3 p_1}{d\lambda^3} + 2^k \frac{d^3p_2}{d\lambda^3}  + 3^k \frac{d^3p_3}{d\lambda^3}   + 4^k \frac{d^3p_4}{d\lambda^3}   \right] \nonumber\\
&+& \dots
\ea
Based on the initial terms of generated sequences it was possible to determine 
%for the cases $k=1$ and $k=2$ are found to be associated with the following algebraic expressions:
\be
\langle n \rangle = \frac{1}{6} \frac{12 + 12 \lambda  + 6 \lambda^2 + \lambda^3}{2 + \lambda}, 
\label{eq:n-1d}
\ee
and 
\be
\langle \Delta n^2 \rangle = \frac{ 180 \lambda  + 360 \lambda^2  + 240 \lambda^3  + 75 \lambda^4   + 12 \lambda^5  +  \lambda^6    }  { 90 (\lambda + 2)^2 },
\label{eq:n2-1d}
\ee
where $\Delta n^2 = n^2 - \langle n \rangle^2$.

\subsection{general asymptotic behavior}

Analytical formulas derived above for $f_A$, $\langle n\rangle$, and $\langle \Delta n^2\rangle$, 
although corresponding to a specific dimension, can be used to infer an asymptotic behavior for 
any dimension using assumptions of the central limit theorem.  Eq. (\ref{eq:fA-1d-R}) for $d=1$ in 
the limit $\lambda\to \infty$ yields 
\be
\frac{f_A}{2} \approx  \frac{1}{\lambda}.  
\label{eq:fA-1d-infty}
\ee
According to the central limit theorem, after a large number of steps (or jumps as applies to our case), 
specific details of a distribution $G$ become unimportant and the sole relevant parameter is the standard 
deviation $\sqrt{\langle \Delta x^2\rangle}$, which based on
Eq. (\ref{eq:SD}) is given by 
\be
\sigma_d    
= \tau v_0 \sqrt{\frac{2}{d}}, 
\label{eq:SD2}
\ee
where the subscript $d$ indicates the dependence on dimension.  
Consequently, a length scale that is specific to a given dimension is $\sigma_d$ (rather than $\tau v_0$ used 
previously).  This means that the wall separation $L$ when partitioned into segments $\sigma_d$ will have
different dimensionless size, 
$$
\lambda_d = \frac{L}{\sigma_d} = \lambda \sqrt{\frac{d}{2}}.  
$$
Based on this, we can redefine the asymptotic formula in Eq. (\ref{eq:fA-1d-infty}) as 
$$
\frac{f_A}{2} \approx  \frac{1}{\sqrt{2}} \frac{1}{\lambda_{d=1}} = \frac{1}{\lambda}.  
$$
Generalizing the above formula to any $d$ yields 
\be
\frac{f_A}{2} \approx   \frac{1}{\sqrt{2}} \frac{1}{\lambda_d} =  \frac{1}{\sqrt{d}} \frac{1}{\lambda}.  
\label{eq:fA-d-infty}
\ee
Following the same procedure, we can generalize the formulas in Eq. (\ref{eq:n-1d}) and Eq. (\ref{eq:n2-1d}), 
%to obtain a general asymptotic behavior for $\langle n \rangle$ and $\langle \Delta n^2 \rangle$, resulting in 
\be
\langle n \rangle \approx \frac{\lambda^2 d}{6}, ~~~~~ \langle \Delta n^2 \rangle \approx  \frac{  \lambda^4  d^2  }  { 90 }.  
\ee

Eq. (\ref{eq:fA-d-infty}) agrees with the asymptotic behavior obtained using alterative derivation in \cite{PRL-Klinger-2022}, 
and which is formulated as 
$$
\frac{f_A}{2} = \pi(0) \approx  \frac{a}{L}, 
$$
where $a$ corresponds to the coefficient of expansion 
%is the coefficient of expansion of the Fourier transformed distribution $G$, 
$
\tilde G(k)  = 1 - a^2 k^2 + O(k^4),
$
where $\tilde G(k)  =   \int_{-\infty}^{\infty} dl\, e^{ik l} G(l) $.

\subsection{the case $d=2$}

Analytical results for $d>1$ are more challenging to derive.  The reason can be traced to the presence of a
logarithmic singularity in $G$.  
For example, the series expansion of $p_n$ in $\lambda$ contains logarithmic terms $\ln\lambda$.  
This makes the computation of $p_n$ from Eq. (\ref{eq:pn}) difficult.  

From the initial terms of the series expansion of $p_n$ we get the initial terms of the expansion of $f_A$, 
\ba
\frac{f_A}{2} &=& \frac{1}{2} -     \lambda \left( \frac{1}{2\pi} - \frac{\gamma  +  \ln \frac{\lambda}{2} } {2\pi} \right)  \nonumber\\
&-&   \lambda^2  \left(  \frac{15  -  \pi^2}{24\pi^2}  -  \frac{ \gamma   +   \ln \frac{\lambda}{2} }{4\pi^2}  \right) \nonumber\\
&+& \dots,
\ea
and the asymptotic behavior in the limit $\lambda\to \infty$ is determined based on Eq. (\ref{eq:fA-d-infty}).  

%$$
%\frac{f_A}{2}    =    \frac{1}{2}    +    \sum_{n=1}^{\infty} a_n\lambda^n    +      \ln\frac{\lambda}{2} \sum_{n=1}^{\infty} b_n\lambda^n   
%$$

\subsection{the case $d=3$}

Following the similar procedure to that for $d=2$, we get the following initial terms for the expansion of $f_A$:  
\ba
\frac{f_A}{2} &=& \frac{1}{2} -    \lambda \left( \frac{1}{4}  -  \frac{ \gamma  +  \ln \lambda }{4}  \right)  \nonumber\\
&-&   \lambda^2  \left(  \frac{27  -  \pi^2}{96}  -  \frac{ \gamma   +   \ln\lambda }{16}  \right) \nonumber\\
&+& \dots
\label{eq:fA-exp-d3}
\ea
The asymptotic behavior in the limit $\lambda\to \infty$ is obtained from Eq. (\ref{eq:fA-d-infty}).

\section{numerical solution}

For $d>1$ where exact results are limited, the integral equation 
can be solved numerically.  To solve Eq. (\ref{eq:pi}), we use an iterative procedure, 
\be
\pi^{(n+1)}(x)   =    \int_{0}^{L} dx'\, \pi^{(n)} (x') G(x'-x)   +   \int_{L}^{\infty} dx'\, G(x'-x), 
\label{eq:iteration}
\ee
with the initial splitting probability corresponding to $d=1$ (see Eq. (\ref{eq:pi-1d})), 
\be
\pi^{(0)}(x)   =    \frac{1}{2}  +   \left( 1 -  f_A^{(0)}  \right) \left( \frac{x}{L}  -  \frac{1}{2} \right), 
\label{eq:pi0}
\ee
where $f_A^{(0)}=\frac{2}{2+\lambda}$ according to Eq. (\ref{eq:fA-1d}).  
%see Eq. (\ref{eq:fA-pi}).  
%Each consecutive iteration $\pi^{(n)}(x)$ converges to $\pi(x)$.  

The first iteration can be carried out analytically and the fraction of adsorbed particles 
for $d=3$ is found to be 
\ba
%&& f_A^{(0)} = \frac{2}{2+\lambda} \nonumber\\
&& f_A^{(1)}   =   1  -   \frac{ \lambda (1 + \lambda\Gamma(0,\lambda) - e^{-\lambda}) }{1+\lambda - e^{-\lambda}}.  
\ea 
The above expression is exact up to the second order term in $\lambda$.  
%$ f_A^{(0)}$ reproduces only the zero order term of the expansion that is exact, and 
%$f_A^{(1)}$ 
%reproduces exactly the zero and first order terms.  
Each subsequent iteration corrects the next higher order term of the expansion of $f_A$.

\begin{figure}[hhhh] 
 \begin{center}
 \begin{tabular}{rrrr}
 \includegraphics[height=0.19\textwidth,width=0.21\textwidth]{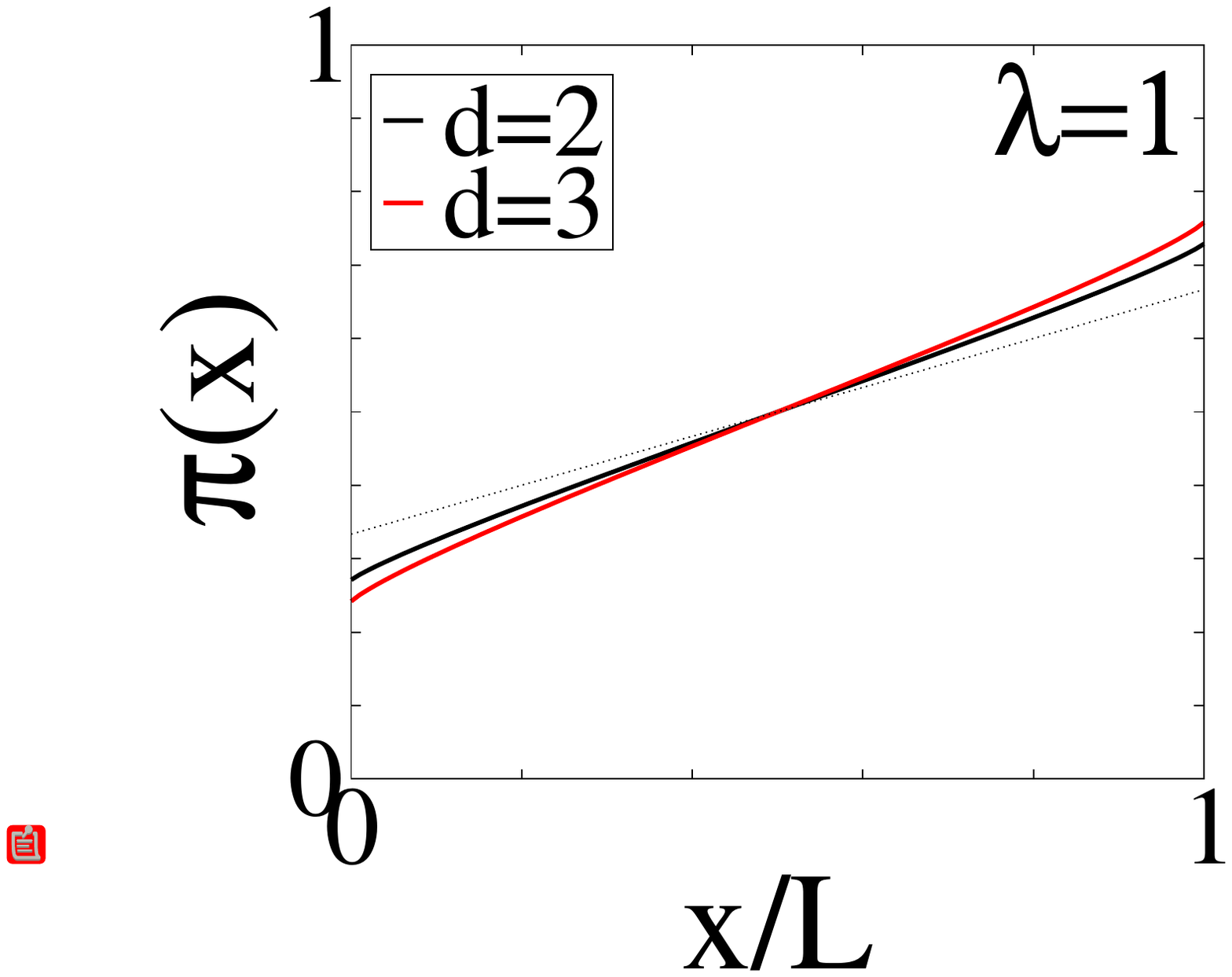} &
 \includegraphics[height=0.19\textwidth,width=0.21\textwidth]{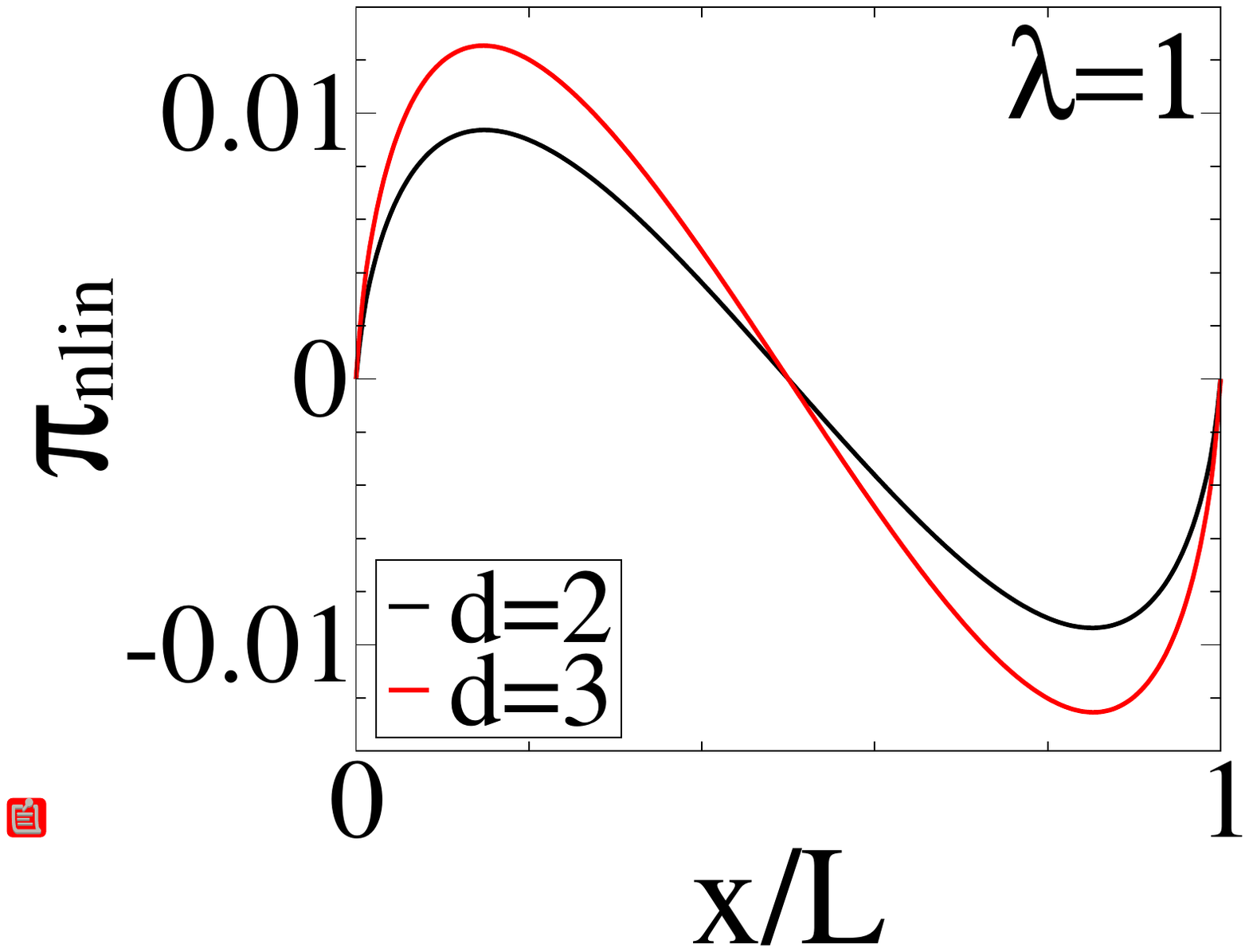} &
\end{tabular}
 \end{center} 
\caption{The splitting probability $\pi(x)$ for $\lambda=1$ and for dimensions $d=2$ and $d=3$.  
A dashed black line represents $\pi(x)$ for $d=1$, given in Eq. (\ref{eq:pi-1d}).
$\pi^{(n)}$ was found to converge after $n=10$ iterations.  
The plots in the figure on the right-hand-side show non-linear contributions of $\pi(x)$.  } 
\label{fig:pi-3d} 
\end{figure}

%It becomes difficult to carry out the analytical iteration beyond $n=1$.  
In Fig. (\ref{fig:pi-3d}) we plot $\pi(x)$ for $\lambda=1$ and for different
dimensions obtained using the numerical iteration in Eq. (\ref{eq:iteration}).  
%The function $\pi^{(n)}$ converges after $n=10$ iterations.  
%%%%%%%%%%%%%%%%%%%%%%
\graphicspath{{figures/}}
%%%%%%%%%%%%%%%%%%%%%%%
To better highlight the difference of each $\pi(x)$, in Fig. (\ref{fig:pi-3d}) (b) we subtract the linear 
term from each curve, $\pi_{nlin} = \pi(x)- \pi_{lin}(x)$, where 
$$
\pi_{lin}(x) = \frac{1}{2}  +   \left( 1-f_A \right) \left( \frac{x}{L}  -  \frac{1}{2} \right).  
$$
Since the stationary distribution of RTP particles is given by $\rho=\pi'(x)$, see Eq. (\ref{eq:rho-pi}), 
from the structure of $\pi_{nlin}$ we can see how divergences at the walls arise.

\section{entropy production, dissipation of heat, pressure}

A particle moving through a dissipating medium with the velocity ${\bf v}$ experiences the frictional 
force ${\bf F}_d = -{\bf v}/\mu$.  An instantaneous rate of the dissipation of heat is given by 
${\bf F}_d\cdot{\bf v} = \dot q = v^2/\mu$.  In the case of an unbounded environment, the motion of 
a RTP particle is unrestricted, thus, the velocity at all times is $|{\bf v}|=v_0$ and the average rate of 
the dissipation of heat (at zero temperature) is 
\be
\langle \dot q \rangle  
%=  \frac{\langle v^2\rangle }{\mu}   
=  \frac{v_0^2}{\mu}.  
\label{eq:dq-T0}
\ee

For RTP particles confined between two walls, walls start to interfere with the motion (resulting in adsorption), 
and so not all of the internal energy is converted to heat.  The fraction of those particles that are not 
adsorbed, $1-f_A$, the heat is dissipated according to Eq. (\ref{eq:dq-T0}).   
But the motion of adsorbed particles is restricted in the direction perpendicular to the walls, thus,  
the dissipation of heat of those particles will be reduced.  In this case, part of the internal energy
that is not converted to heat is manifested as pressure generated by adsorbed particles
pushing against the walls. 

%For example, in the case that an adsorbed particle stops moving, the heat stops being produced.  The 
%total internal energy in this case is converted to work, manifested as pressure as adsorbed particles
%push against a wall. 

%Thus, the internal energy of a particle in confinement can be converted to both heat and work. 
%The heat that is produced is related to the entropy production rate $\Pi$
%via the relation $T\Pi = \langle \dot q \rangle$ \cite{PRE-Frydel-2023a}, and the 
%work that is produced by adsorbed particles is manifested as pressure.  

For a one-dimensional system things are simple.  Adsorbed particles are motionless 
and the heat is dissipated only by particles that are free,  
\be
\langle \dot q\rangle  =  \frac{v_0^2}{\mu}(1-f_A).  
\label{eq:dq-1d}
\ee
The force exerted on a single wall by adsorbed particles is given by 
\be
F_p = \frac{v_0}{\mu} \frac{f_A}{2}.  
\label{eq:Fp-1d}
\ee
%Considering contributions of the two walls, the rate with which work is produced is
%$\langle \dot w\rangle   =  2 F_p v_0$, which results in
%\be
%\langle \dot w\rangle  =  \frac{v_0^2}{\mu} f_A.  
%\ee
%The sum $\langle \dot q\rangle + \langle \dot w\rangle$ is conserved and correspond to the 
%rate with which the internal energy of a particle is expanded.  

%$$
%F_p^{(L)} =  \frac{1}{\tau \mu}  \int_{L}^{\infty} dx'\, x' \, \rho(x') G(x'-x)   +  \frac{1}{\tau \mu}  \frac{f_A}{2} \int_L^{\infty} dx\, x \, G(x-L)
%$$
%$$
%F_p^{(0)} =  -\frac{1}{\tau \mu}  \int_{-\infty}^{0} dx'\, x' \, \rho(x') G(x'-x)   -  \frac{1}{\tau \mu}  \frac{f_A}{2} \int_{-\infty}^{0} dx\, x \, G(x-L)
%$$

For higher dimensions things are more complicated as the 
%motion of adsorbed particles is restricted but not eliminated.  The 
motion is restricted only in the direction perpendicular to the walls and 
particles remain free to move in parallel directions.  Thus, to calculate the dissipation of heat, we need to take 
into account the motion along the wall plane.  

%In Langevin simulation we would calculate the $F_p$ by calculating the velocity component 
%perpendicular to the wall, which corresponds to $v_x\equiv v$.  and the force is 
%$F_p = \langle v \rangle_{c} /\mu$.  The question is, can we calculate $\langle v \rangle_{c}$
%from the jump-process framework.  The initial answer is no, since within that framework
%we only know the length of jumps and have no information about the velocity.  

In our description we only consider the velocity component perpendicular to the wall, 
$v \equiv v_{\perp}$, such that $|v|\leq v_0$.   If $\langle \dots \rangle_A$ denotes an average 
quantity calculated by considering only adsorbed particles, then the   
%which is the part of a velocity that is quenched and $v_{\perp}^2 + v_{\parallel}^2 = v_0^2$, then the rate 
dissipation of heat can be defined as 
\be
\langle \dot q \rangle =  \frac{v_0^2}{\mu}   -   \frac{\langle v^2 \rangle_A}{\mu} f_A,
\label{eq:dq-gen}
\ee
where the first term is the dissipation of heat of particles in an unbounded space, 
and the second term subtracts the heat that is not dissipated due to 
adsorbed particles with restricted motion.  

To calculate $\langle \dot q \rangle$, we need to obtain an expression for $\langle v^2 \rangle_A$. 
Such an expression, within the splitting probability framework, for $\langle v^n \rangle_{A}$ is given by 
%First we note that the probability that a particle, initially at the location $x=x_0$, 
%will become adsorbed is $\int_L^{\infty} dx\, G(x-x_0)$.  We can also relate the average velocity of 
%an adsorbed particle by the length by which it crossed the wall, $\sim \int_L^{\infty} dx\, (x-L) G(x-x_0)$.   
%However, since the size of a jump depends on the product of two independent random variables $v$ and $b$, 
%we need still do divide the above expression by $\langle t_p\rangle=\tau$, leading to 
%$\sim \int_L^{\infty} dx\, (x-L) G(x-x_0)/\tau$.  
%Based on this, we propose the following expression for $\langle v^n \rangle_{A}$,   
%\ba
%\langle v_{\perp}^n \rangle_{a}  f_A  &=&  \frac{2}{n!\tau^n} \int_0^L dx'\, \rho(x')  \int_{L}^{\infty} dx\, (x - L)^n \, G(x'-x)   \nonumber\\ 
%&+&   \frac{2}{n! \tau^n} \int_L^{\infty} dx\, (x - L)^n \,  \left[ \frac{f_A}{2} G(x-L)   +   \frac{f_A}{2} G(x) \right]. \nonumber\\
%%+   \frac{f_A}{\tau\mu} \int_L^{\infty} dx\, x \, G(x) 
%\label{eq:vw}
%\ea
\ba
\langle v^n \rangle_{A}&=&  \frac{2}{n!\tau^n f_A} \int_0^L dx'\, \rho(x')  \int_{L}^{\infty} dx\, (x - L)^n \, G(x'-x)   \nonumber\\ 
&+&   \frac{1}{n! \tau^n} \int_L^{\infty} dx\, (x - L)^n \,  \left[ G(x-L)   +   G(x) \right], %\nonumber\\
\label{eq:vw}
\ea
where we used $\langle t_p^n\rangle = n! \tau^n$.  The second line represents the probability that 
a particle becomes re-adsorbed.  
Eq. (\ref{eq:vw}) can be verified for $d=1$, in which case $G(x)$ is an exponential 
function and $\langle v^n \rangle_{A} = v_0^n$.  For $d=2$ and $d=3$, the expression is 
verified by simulations.  

From Eq. (\ref{eq:dq-gen}) and Eq. (\ref{eq:vw}) (and using Eq. (\ref{eq:rho-pi}) and 
Eq. (\ref{eq:fA-pi})), we get an expression for the dissipation of heat in terms of a splitting probability, 
%\ba
%\langle \dot q \rangle   &=&   \frac{v_0^2}{\mu}   -   \frac{1}{\mu \tau^2 }  \int_{0}^{\infty} dx\, x^2 \, G(x)   \nonumber\\ 
%  &+&   \frac{2}{\mu \tau^2 } \int_0^L dx'\, \pi(x')  \int_{L}^{\infty} dx\, (x - L) \, G(x-x').  
%\label{eq:dq-pi}
%\ea
\be
\langle \dot q \rangle   =   \frac{v_0^2}{\mu}   \frac{d-1}{d} +   \frac{2}{\mu \tau^2 } \int_0^L dx'\, \pi(x')  \int_{L}^{\infty} dx\, (x - L) \, G(x-x').  
\label{eq:dq-pi}
\ee
For $d=1$ the equation above recovers the result in Eq. (\ref{eq:dq-1d}).  

It is also possible to obtain an expression for the force acting on a single wall, for a general dimension defined as 
\be
F_p = \frac{f_A}{2} \frac{\langle v \rangle_A}{\mu}.  
\label{eq:Fp-gen}
\ee
Using Eq. (\ref{eq:vw}), together with Eq. (\ref{eq:rho-pi}) and Eq. (\ref{eq:fA-pi}), we get 
\be
F_p  =    \frac{1}{\tau \mu}  
% \frac{\Gamma(d/2)}{\Gamma(1/2)\Gamma(d/2+1/2)}
\int_{0}^{\infty} dx\, x \, G(x)   
-   \frac{1}{\tau \mu } \int_0^L dx'\, \pi(x')  \int_{L}^{\infty} dx\, G(x'-x) .  
%&+&  \frac{1}{\tau \mu }   \int_{L}^{\infty} dx\, (x - L) \, G(x-L).  
\label{eq:Fp}
\ee
Once again, we can verify this expression for $d=1$, which recovers the result in Eq. (\ref{eq:Fp-1d}).

\begin{figure}[hhhh] 
 \begin{center}
 \begin{tabular}{rrrr}
 \includegraphics[height=0.19\textwidth,width=0.21\textwidth]{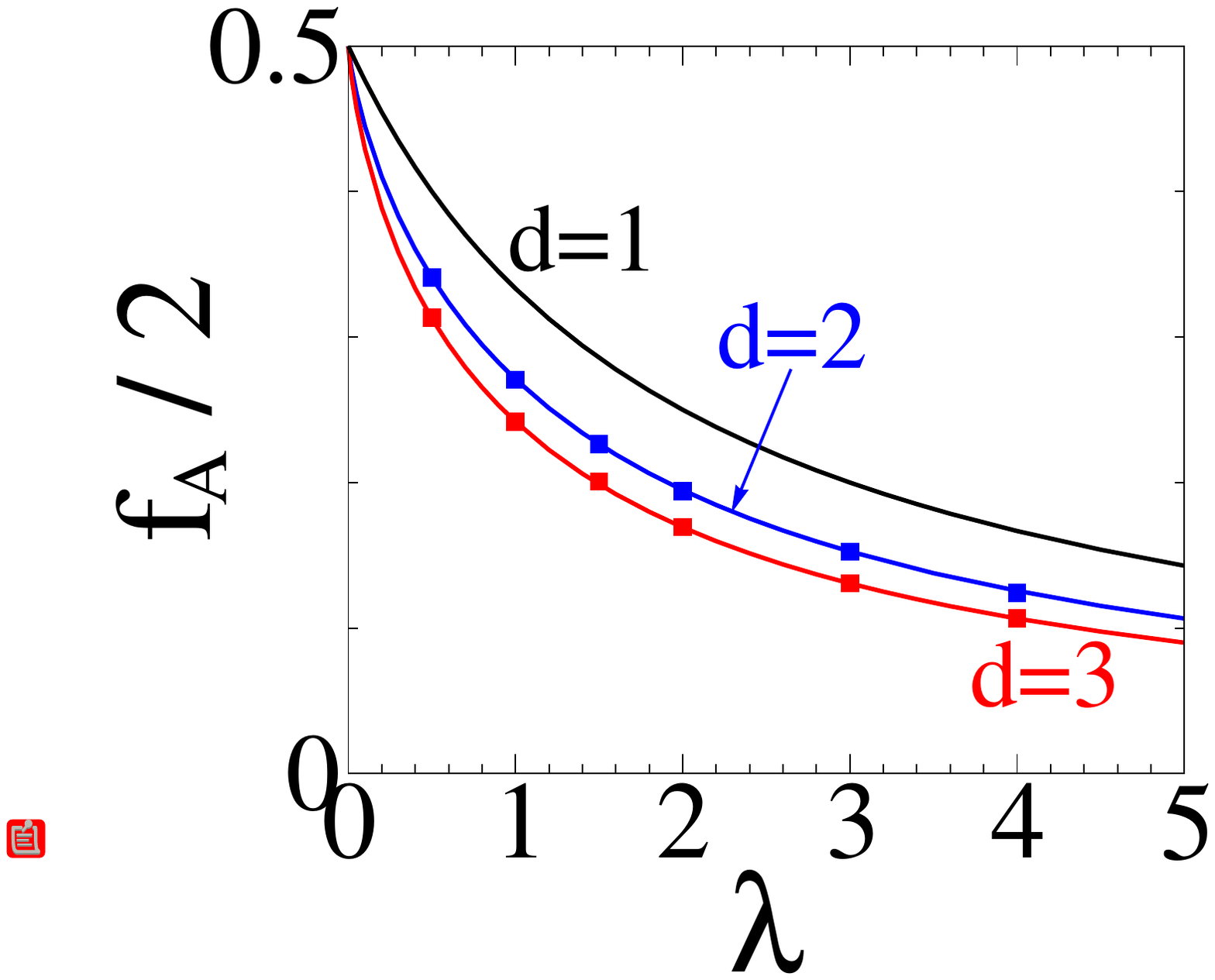} &
 \includegraphics[height=0.19\textwidth,width=0.21\textwidth]{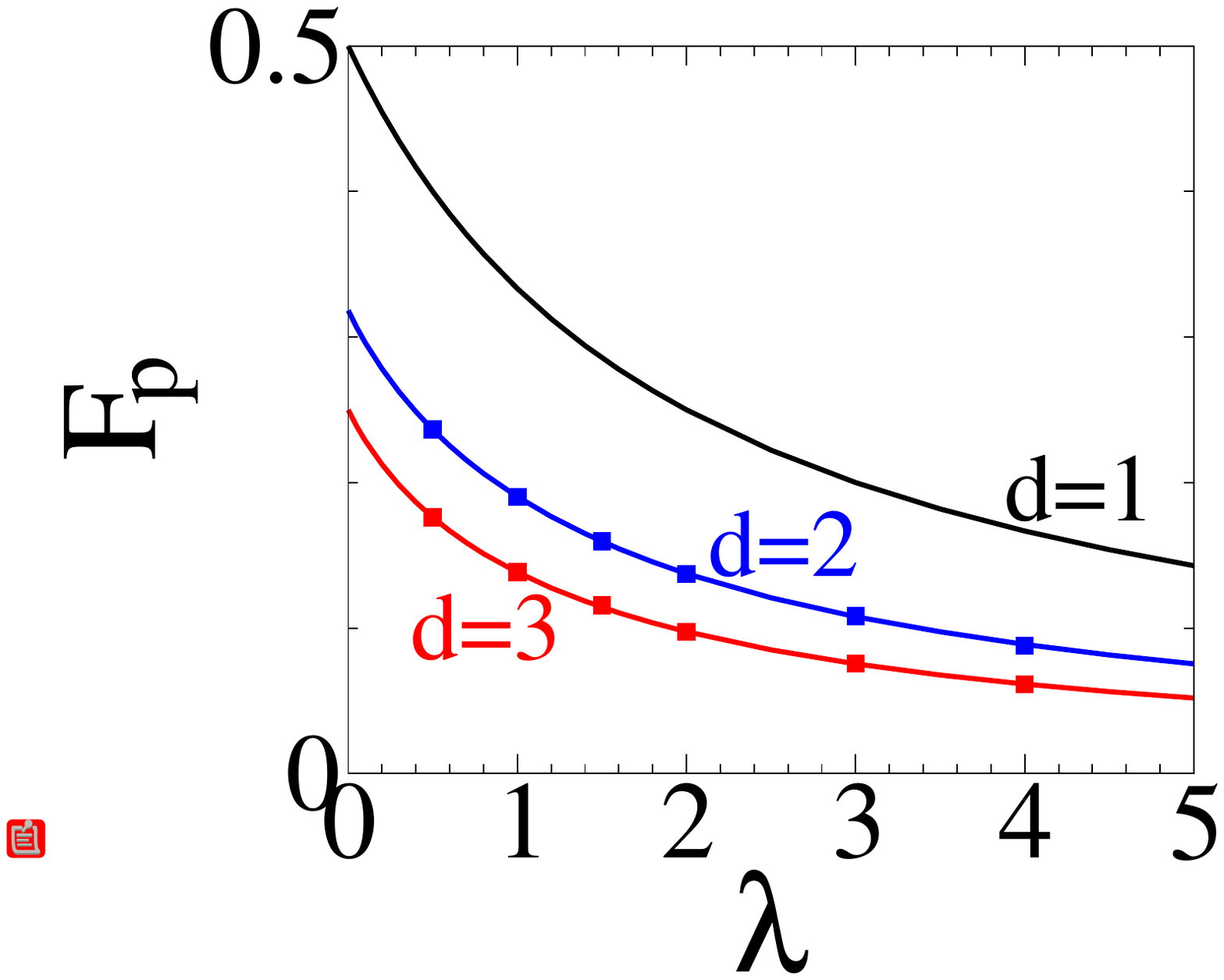} &
\end{tabular}
 \end{center} 
\caption{The fraction of adsorbed particles on a single wall, $f_A/2=\pi(0)$, and the 
force exerted by those particles on a single wall, $F_p=f_A\langle v\rangle_A/(2\mu)$, 
as a function of $\lambda$.  In the second figure it is assumed that $v_0=\mu=1$.  
The square symbols represent the data points of a dynamic simulation to confirm the numerical 
results of the splitting probability framework.} 
\label{fig:fA-3d} 
\end{figure}

In Fig. (\ref{fig:fA-3d}) we plot the fraction particles adsorbed onto a single wall, $f_A/2=\pi(0)$, 
and the force exerted on a wall by those particles, $F_p$.  
The results are obtained from $\pi(x)$ calculated numerically and then using the respective
formulas.  We compare all the results with data points from a simulation based on the Langevin equation.  
%%%%%%%%%%%%%%%%%%%%%%
\graphicspath{{figures/}}
%%%%%%%%%%%%%%%%%%%%%%%

The figure shows that adsorption decreases with increasing system dimension.  
%This behavior
%can be traced to the probability distribution of jumps $G$ that are less spread out as 
%the system dimension increases, see Eq. (\ref{eq:SD}).
%This was already demonstrated from the asymptotic formula in Eq. (\ref{eq:fA-d-infty}).  
The same trend is observed for the force $F_p$.  
At $\lambda=0$, that is, at the point where the two walls touch one another 
and all particles are adsorbed, the exerted force increases with decreasing $d$.  
This happens because 
%the average velocity 
%of adsorbed particles $\langle v \rangle_A$ decreases with increasing $d$, 
%since the probability of 
for higher dimensions the probability of motion in the direction parallel to the walls increases.

\begin{figure}[hhhh] 
 \begin{center}
 \begin{tabular}{rrrr}
 \includegraphics[height=0.19\textwidth,width=0.21\textwidth]{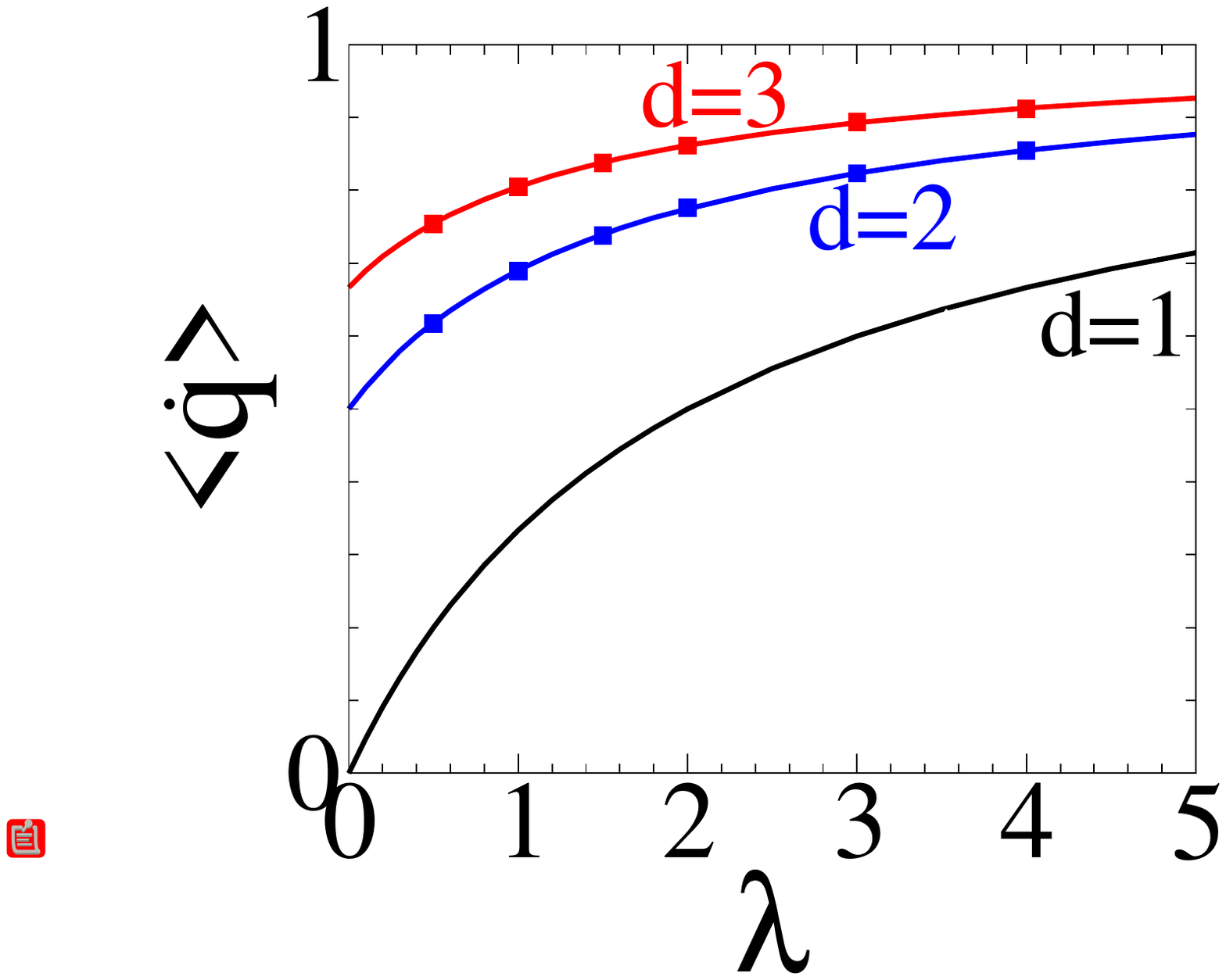} &
\end{tabular}
 \end{center} 
\caption{Dissipation of heat calculated numerically from Eq. (\ref{eq:dq-pi}).  
It is assumed that $v_0=\mu=1$. The square symbols represent the data
points of a dynamic simulation based on the Langevin equation.  } 
\label{fig:dq-3d} 
\end{figure}

In Fig. (\ref{fig:dq-3d}) we plot the dissipation of heat (related to the entropy production 
rate as $T\Pi = \langle\dot q\rangle$) as a function of $\lambda$.  
%%%%%%%%%%%%%%%%%%%%%%
\graphicspath{{figures/}}
%%%%%%%%%%%%%%%%%%%%%%%
The dissipation of heat increases with increasing $d$.  
This again has two causes.  The first cause can be traced to the fact that 
%the dissipation of heat decreases
%for adsorbed particles (since part of an internal energy is used to generate pressure), and 
for a lower dimensional system the adsorption is greater, see Fig. (\ref{fig:fA-3d}), 
and adsorbed particle dissipate less heat.  
The second cause is that for a larger $d$, the motion of 
adsorbed particles is less restricted as there are more available directions 
along the wall plane.  Note that for $\lambda=0$, where all particles are adsorbed, 
the dissipation of heat vanishes for $d=1$.  Adsorbed particles in this case 
are motionless as no parallel motion along the wall is available.  
In this case all the internal energy is used to generate pressure.

The dissipation of heat for different dimensions can be accurately fit into the following ansatz, 
\be
\langle \dot q\rangle  =   \frac{v_0^2}{\mu} \left(  \frac{d-1}{d}   +  \frac{1}{d} \frac{\lambda}{c_d + \lambda} \right),
\ee
with fitting coefficients $c_1=2$, $c_2 = 1.63$, $c_3 = 1.4106$ (in the case $d=1$ the ansatz 
corresponds to the exact formula).  
The exact expansion up to the first order term is
\be
\langle \dot q\rangle  =   \frac{v_0^2}{\mu} \frac{d-1}{d}   +  a_d \lambda + \dots 
\ee
with
$$
a_d = \frac{1}{2} \frac{\Gamma(d/2) }{\Gamma(1/2) \Gamma(d/2 + 1/2) }.
$$

\section{Conclusion}
\label{sec:5}

The minimal system of RTP particles confined between two walls, despite its apparent 
simplicity, has not been satisfactorily treated for two- and three-dimensional cases.  In this work 
we treat this system by changing theoretical framework from the Fokker-Planck description to a 
framework that is consistent with the problem of splitting probability and represented by the 
integral equation in Eq. (\ref{eq:pi}) together with the relations in Eq. (\ref{eq:rho-pi}) and Eq. (\ref{eq:fA-pi}).  

The alternative framework is obtained by modifying the microscopic motion from continuous time
dynamics to jump process, in such a way that macroscopic stationary properties of 
both processes are identical.  To do this, it was necessary to determine a correct probability 
distribution of discontinuous jumps $G(\Delta x)$ that is specific to each dimension, see Eq. (\ref{eq:G-0}). 
For $d=1$, $G(\Delta x)$ is an exponential function and for the cases $d=2$ and $d=3$, $G(\Delta x)$
has a logarithmic singularity at $\Delta x=0$.   

Based on the integral equation it was possible to identify the divergence in the stationary distribution $\rho(x)$
with the probability distributions $G$ positionat at both walls, and 
representing the distribution of particles recently de-adsorbed.   

The integral equation can be solved exactly for the case $d=1$.  
The exact treatment of the integral equation for $d=2$ and $d=3$ is inhibited by the occurrence of the 
logarithmic singularity in $G$.  Some limited exact results are possible, such as asymptotic 
behavior and the initial terms of the expansion of various physical quantities, such as the fraction 
of adsorbed particles or the dissipation of heat.  
To solve the integral equation for $\pi(x)$ for those cases, we resort to a numerical iterative 
scheme.

The procedure to substitute a microscopic process based on continuous dynamics with another 
process based on discontinuous jumps is not limited or specific to a confinement between walls.  
At least in principle, it is valid for any other confinement or external potential.  The resulting jump
algorithm could be a valid alternative to simulations based on the Langevin equation, 
as long as the focus is on stationary properties.  
%our interest is limited to stationary properties.  In this light, jump algorithm deserves further attention.  
%could serve as a bas
%when simulating RTP particles in any
%external potential, it could be used as a valid simulation algorithm alternative to dynamical
%simulations --- that is, our interest is limited to stationary properties.  These ideas are 
%worth of further examination.  

%The transformation was not carried directly from the Fokker-Planck description in Eq. (\ref{eq:FP}) 
%but by modifying the microscopic process that substitutes the Langevin dynamics 
%with the jump-process algorithm, governed by the probability distributions of jump lengths
%that depend on a system dimension and given in Eq. (\ref{eq:G-0}).

\appendix
\section{Jump-algorithm simulation}

In this section we compare two simple Python codes for sampling configurations in the inteval
$x\in[0,L]$.  The simulations are specific to $d=3$, for which the distribution of velocities
$v$ is uniform on the interval $[-v_0,v_0]$, see Eq. (\ref{eq:pv}).

The simulation below is based on the numerical integration of the Langevin equation, see 
Eq. (\ref{eq:langevin-0}). 
Each step in the simulation corresponds to the next discrete time step $t+dt$.   
Physical parameters are v0$=v_0$, tau$=\tau$, and L$=L$.  A new velocity $v$ is 
selected from the uniform distribution at the end of a persistence time tp$=t_p$.  
A particle is considered as adsorbed when it crosses the location of a wall at 
$x=0$ and $x=L$.  Once a particle crosses a wall, it is brought to the location of a
wall it just crossed.  
%\vspace{0.5cm}\noindent 
%\begin{figure}
\begin{lstlisting}%[basicstyle=\small]

import math
import random
import numpy as np

tau=1 #physical parameters
v0=1
L=1

dt = 0.001    #discrete time interval
NSTEP  = 100000 #number of time steps

time=0
tp = np.random.exponential(tau)
v = v0*np.random.uniform(-1,1)
x=0   #initial position 
for i in range(0,NSTEP):
        time += dt
        if time>tp:   #new velocity selected 
                tp = np.random.exponential(tau)
                time=0
                v = v0*np.random.uniform(-1,1)
        x += v*dt
        if x>L:    #adsorption onto a wall at x=L
                x=L   
        elif x<0:  #adsorption onto a wall at x=0
                x=0 

\end{lstlisting}

The Python code below is for a jump-process algorithm.  Note that unlike the 
code above, there is no time variable.  Each step corresponds to the next jump.  
Note that the jump-process algorithm is simpler and more efficiently samples
different configurations.  
%\vspace{0.5cm}\noindent 
%\begin{figure}
\begin{lstlisting}%[basicstyle=\small]

import math
import random
import numpy as np

tau=1 
v0=1  
L=1   

NSTEP=10000
x=0
for i in range(0,NSTEP):
        v = v0*np.random.uniform(-1,1)
        tp = np.random.exponential(tau)
        x += tp*v     #particle jump 
        if x>L:    
        		x=L  
        elif x<0:  
        		x=0
                
\end{lstlisting}

%------------------------------------------------
\begin{acknowledgments}
D.F. acknowledges financial support from FONDECYT through grant number 1241694.  
\end{acknowledgments}

\section{DATA AVAILABILITY}
The data that support the findings of this study are available from the corresponding author upon 
reasonable request.

\bibliography{RTP-wall}

\end{document}